\documentclass{article}
\usepackage{amssymb}
\usepackage{amsmath}
\usepackage{makeidx}

\newtheorem{theorem}{Theorem}

\newtheorem{definition}[theorem]{Definition}

\newtheorem{lemma}[theorem]{Lemma}

\newtheorem{proposition}[theorem]{Proposition}
\newtheorem{remark}[theorem]{Remark}

\input{tcilatex}
\begin{document}

\title{Hylomorphic solitons and charged Q-balls: existence and stability}
\author{Vieri Benci\thanks{
Dipartimento di Matematica Applicata, Universit\`a degli Studi di Pisa, Via
F. Buonarroti 1/c, Pisa, ITALY and Department of Mathematics, College of
Science, King Saud University, Riyadh, 11451, SAUDI ARABIA. e-mail: \texttt{%
benci@dma.unipi.it}} \and Donato Fortunato\thanks{%
Dipartimento di Matematica, Universit\`{a} degli Studi di Bari, Via Orabona,
Bari, ITALY. e-mail: \texttt{\ fortunat@dm.uniba.it, }}}
\maketitle

\begin{abstract}
In this paper we give an abstract definition of solitary wave and soliton
and we develope an abstract existence theory. This theory provides a
powerful tool to study the existence of solitons for the Klein-Gordon
equations as well as for gauge theories. Applying this theory, we prove the
existence of a continuous family of stable charged Q-balls.
\end{abstract}

\tableofcontents

\ \bigskip

\bigskip


\bigskip

\bigskip

\section{Introduction}

Loosely speaking a solitary wave is a solution of a field equation whose
energy travels as a localized packet and which preserves this localization
in time. A \textit{soliton} is a solitary wave which exhibits some strong
form of stability so that it has a particle-like behavior (see e.g. \cite%
{Rajaraman}, \cite{yangL}, \cite{rub} and the references therein contained).
We are interested in a class of solitons which, following \cite{benci}, \cite%
{BBBM}, \cite{bbbs}, \cite{befoni}, we call hylomorphic. Their existence is
due to an interplay between energy and charge. These solitons include the $Q$%
-balls, which are spherically symmetric solutions of the nonlinear
Klein-Gordon equation and which have been studied since the pioneering
papers \cite{rosen68} and \cite{Coleman86}. Q-balls arise in a theory of
bosonic particles (see \cite{kus}, \cite{leepa}), when there is an
attraction between the particles. Roughly speaking, a Q-ball is a
finite-sized "bubble" containing a large number of particles. The Q-ball is
stable against fission into smaller Q-balls since, due to the attractive
interaction, the Q-ball is the lowest-energy configuration of that number of
particles. Q-balls also play an important role in the study of the origin of
the matter that fills the universe (see \cite{dod}).

In this paper we give an abstract definition of solitary wave and soliton
and we develope an abstract existence theory. This theory provides a
powerful tool to study the existence of solitons for the Klein-Gordon
equations as well as for gauge theories (see \cite{bebo12}). Most of the
existence results in the present literature can be deduced in the framework
of this theory using Th.\ref{astra1} or \ref{astra1+}, as it is shown in a
forthcoming book \cite{befolib}. We get a new result applying Th.\ref%
{astra1+} to the study of \textit{charged Q-balls}. Let us describe this
result.

If the Klein-Gordon equations are coupled with the Maxwell equations (NKGM),
then the relative solitary waves are called \textit{charged}, or \textit{%
gauged Q-balls} (see e.g.\cite{Coleman86}). The existence of charged Q-balls
is stated in \cite{befo}, \cite{befo08}, \cite{befospin}, \cite{mug}.
However, in these papers there are not stability results and hence the
existence of solitons for NKGM (namely \textit{stable} charged Q-balls) was
an open question.

The problem with the stability of charged Q-balls is that the electric
charge tends to brake them since charges of the same sign repel each other.
In this respect Coleman, in his celebrated paper \cite{Coleman86}, says
\textquotedblright \textit{I have been unable to construct Q-balls when the
continuous symmetry is gauged. I think what is happening physically is that
the long-range force caused by the gauge field forces the charge inside the
Q-ball to migrate to the surface, and this destabilizes the system, but I am
not sure of this}\textquotedblright .

A partial answer to this question is in \cite{befo11max} where the existence
of stable charged Q-balls is established provided that the interaction
between matter and gauge field is sufficiently small. Theorem \ref{astra1+}
allows to extend this result and to prove the existence of a continuous
family of stable charged Q-balls. More precisely, we prove that there is a
family of Q-balls $\left\{ \mathbf{u}_{\delta }\right\} _{\delta \in \left(
0,\delta _{\infty }\right) }$ whose energy and charge are decreasing with $%
\delta .$

\section{Solitary waves and solitons}

In this section we construct a functional abstract framework which allows to
define solitary waves, solitons and hylomorphic solitons.

\subsection{Definitions of solitons\label{be}}

\textit{Solitary waves} and solitons are particular \textit{states} of a
dynamical system described by one or more partial differential equations.
Thus, we assume that the states of this system are described by one or more 
\textit{fields} which mathematically are represented by functions 
\begin{equation*}
\mathbf{u}:\mathbb{R}^{N}\rightarrow V
\end{equation*}%
where $V$ is a vector space with norm $\left\vert \ \cdot \ \right\vert _{V}$
and which is called the internal parameters space. We assume the system to
be deterministic; this means that it can be described as a dynamical system $%
\left( X,\gamma \right) $ where $X$ is the set of the states and $\gamma :%
\mathbb{R}\times X\rightarrow X$ is the time evolution map. If $\mathbf{u}%
_{0}(x)\in X,$ the evolution of the system will be described by the function 
\begin{equation}
\mathbf{u}\left( t,x\right) :=\gamma _{t}\mathbf{u}_{0}(x).  \label{flusso}
\end{equation}%
We assume that the states of $X$ have "finite energy" so that they decay at $%
\infty $ sufficiently fast and that%
\begin{equation}
X\subset L_{loc}^{1}\left( \mathbb{R}^{N},V\right) .  \label{lilla}
\end{equation}

Thus we are lead to give the following definition:\label{pag}

\begin{definition}
\label{ft}A dynamical system $\left( X,\gamma \right) $ is called of FT type
(field-theory-type) if $X$ is a Hilbert space of fuctions of type (\ref%
{lilla}).
\end{definition}

For every $\tau \in \mathbb{R}^{N},$ and $\mathbf{u}\in X$, we set 
\begin{equation}
\left( T_{\tau }\mathbf{u}\right) \left( x\right) =\mathbf{u}\left( x+\tau
\right) .  \label{ggg}
\end{equation}%
Clearly, the group 
\begin{equation}
\mathcal{T}=\left\{ T_{\tau }|\ \tau \in \mathbb{R}^{N}\right\}  \label{gg}
\end{equation}%
is a unitary representation of the group of translations.

\begin{definition}
\label{carla}A set $\Gamma \subset X$ is called compact up to space
tanslations or $\mathcal{T}$-compact if for any sequence $\mathbf{u}%
_{n}(x)\in \Gamma \ $there is a subsequence $\mathbf{u}_{n_{k}}$ and a
sequence $\tau _{k}\in \mathbb{R}^{N}$ such that $\mathbf{u}_{n_{k}}(x-\tau
_{k})$ is convergent.
\end{definition}

Now, we want to give a very abstract definition of solitary wave. As we told
in the introduction, a solitary wave is a field whose energy travels as a
localized packet and which preserves this localization in time. For example,
consider a solution of a field equation having the following form: 
\begin{equation}
\mathbf{u}\left( t,x\right) =u_{0}(x-vt-x_{0})e^{i(v\mathbf{\cdot }x\mathbf{-%
}\omega t)};\ u_{0}\in L^{2}(\mathbb{R}^{N});  \label{solwav}
\end{equation}%
for every $x_{0},v\in \mathbb{R}^{N},\omega \in \mathbb{R},$ $\mathbf{u}%
\left( t,x\right) $ is a solitary wave. The evolution of a solitary wave is
a translation plus a mild change of the internal parameters (in this case
the phase).

This situation can be formalized by the following definition:

\begin{definition}
\label{solw} If $\mathbf{u}_{0}\in X,$ we define the closure of the orbit of 
$\mathbf{u}_{0}$ as follows:%
\begin{equation*}
\mathcal{O}\left( \mathbf{u}_{0}\right) :=\overline{\left\{ \gamma _{t}%
\mathbf{u}_{0}(x)\ |\ t\in \mathbb{R}\right\} }.
\end{equation*}%
A state $\mathbf{u}_{0}\in X$ is called solitary wave if

\begin{itemize}
\item (i) $0\notin \mathcal{O}\left( \mathbf{u}_{0}\right) ;$

\item (ii) $\mathcal{O}\left( \mathbf{u}_{0}\right) $ is $\mathcal{T}$%
-compact.
\end{itemize}
\end{definition}

Clearly, (\ref{solwav}) describes a solitary wave according to the
definition above. The standing waves, namely objects of the form 
\begin{equation}
\gamma _{t}\mathbf{u=u}(t,x)=u(x)e^{-i\omega t},\ \ u\in L^{2}(\mathbb{R}%
^{N}),\ u\neq 0,  \label{stand}
\end{equation}%
probably are the "simplest" solitary waves. In this case the orbit $\mathcal{%
O}\left( \mathbf{u}_{0}\right) $ itself is compact.

Take $X=L^{2}(\mathbb{R}^{N})$\ and\ $u\in X;$ if $\gamma _{t}u=u\left( 
\frac{x}{e^{t}}\right) $, $u$ is not a solitary wave, since (i) of the above
definition is violated; if $\gamma _{t}u=\frac{1}{e^{t}}u\left( \frac{x}{%
e^{t}}\right) ,$ $u$ is not a solitary wave since (ii) of Def. \ref{solw}
does not hold. Also, according to our definition, a "couple" of solitary
waves is not a solitary wave: for example 
\begin{equation*}
\gamma _{t}\mathbf{u}=\left[ u_{0}(x-vt)+u_{0}(x+vt)\right] e^{i(v\mathbf{%
\cdot }x\mathbf{-}\omega t)},
\end{equation*}%
is not a solitary wave since (ii) is violated.

The \textit{solitons%
\index{soliton}} are solitary waves characterized by some form of stability.
To define them at this level of abstractness, we need to recall some well
known notions in the theory of dynamical systems.

\begin{definition}
A set $\Gamma \subset X$ is called \textit{invariant} if $\forall \mathbf{u}%
\in \Gamma ,\forall t\in \mathbb{R},\ \gamma _{t}\mathbf{u}\in \Gamma .$
\end{definition}

\begin{definition}
Let $\left( X,d\right) $ be a metric space and let $\left( X,\gamma \right) $
be a dynamical system. An invariant set $\Gamma \subset X$ is called stable,
if $\forall \varepsilon >0,$ $\exists \delta >0,\;\forall \mathbf{u}\in X$, 
\begin{equation*}
d(\mathbf{u},\Gamma )\leq \delta ,
\end{equation*}%
implies that 
\begin{equation*}
\forall t\geq 0,%
\text{ }d(\gamma _{t}\mathbf{u,}\Gamma )\leq \varepsilon .
\end{equation*}
\end{definition}

Now we are ready to give the definition of soliton:

\begin{definition}
\label{ds} A state $\mathbf{u}\in X$ is called soliton if $\mathbf{u}\in
\Gamma \subset X$ where

\begin{itemize}
\item (i) $\Gamma $ is an invariant stable set,

\item (ii) $\Gamma $ is $\mathcal{T}$-compact

\item (iii) $0\notin \Gamma $.
\end{itemize}
\end{definition}

The above definition needs some explanation. First of all notice that every $%
\mathbf{u}\in \Gamma $ is a soliton and that every soliton is a solitary
wave. Now for simplicity, we assume that $\Gamma $ is a manifold (actually,
in many concrete models, this is the generic case). Then (ii) implies that $%
\Gamma $ is finite dimensional. Since $\Gamma $ is invariant, $\mathbf{u}%
_{0}\in \Gamma \Rightarrow \gamma _{t}\mathbf{u}_{0}\in \Gamma $ for every
time. Thus, since $\Gamma $ is finite dimensional, the evolution of $\mathbf{%
u}_{0}$ is described by a finite number of parameters$.$ The dynamical
system $\left( \Gamma ,\gamma \right) $\ behaves as a point in a finite
dimensional phase space. By the stability of $\Gamma $, a small perturbation
of $\mathbf{u}_{0}$ remains close to $\Gamma .$ However, in this case, its
evolution depends on an infinite number of parameters. Thus, this system
appears as a finite dimensional system with a small perturbation.

\bigskip

\textbf{Example\label{ese}. }We will illustrate the definition \ref{ds} with
an example. Consider the solitary wave (\ref{solwav}) and the set%
\begin{equation*}
\Gamma _{v}=\left\{ u(x-x_{0})e^{i\left( v\mathbf{\cdot }x-\theta \right)
}\in H^{1}\left( \mathbb{R}^{N},\mathbb{C}\right) :x_{0}\in \mathbb{R}^{N};\
\theta \in \mathbb{R}\right\} .
\end{equation*}%
(\ref{solwav}) is a soliton provided that $\Gamma _{v}$ is stable; in fact
the following conditions are satisfied:

\begin{itemize}
\item The dynamics on $\Gamma _{v}$ is given by the following equation:%
\begin{equation*}
\gamma _{t}\left[ u(x-x_{0})e^{i\left( v\mathbf{\cdot }x-\theta \right) }%
\right] =u(x-vt-x_{0})e^{i(v\mathbf{\cdot }x\mathbf{-}\theta \mathbf{-}%
\omega t)}.
\end{equation*}%
This dynamics implies that $\Gamma _{v}$ is invariant and that (iii) holds.

\item we have assumed that $\Gamma _{v}$ is stable; in this case any
perturbation of our soliton has the following structure:%
\begin{equation*}
u(t,x)=u(x-vt-x_{0}(t))e^{i\left( v\mathbf{\cdot }x-\theta (t)\right)
}+w(t,x)
\end{equation*}%
where $x_{0}(t),$ $\theta (t)$ are suitable functions and $w(t,x)\ $is a
perturbation small in $H^{1}\left( \mathbb{R}^{N},\mathbb{C}\right) .$

\item $\Gamma _{v}$ is $\mathcal{T}$-compact; actually it is isomorphic to $%
\mathbb{R}^{N}\times S^{1}$.
\end{itemize}

\subsection{Definition of hylomorphic solitons\label{util}}

We now assume that the dynamical system $\left( X,\gamma \right) $ has two
constants of motion: the energy $E$ and the charge $C.$ At the level of
abstractness of this section (and the next one), the name energy and charge
are conventional, but in our applications,\ $E$ and $C$ will be the energy
and the charge as defined in section \ref{cl}.

\begin{definition}
\label{tdc}A solitary wave $\mathbf{u}_{0}\in X$ is called \textbf{standing} 
\textbf{hylomorphic soliton%
\index{soliton!hylomorphic}} if it is a soliton according to Def. \ref{ds}
and if $\Gamma $ has the following structure%
\begin{equation}
\Gamma =\Gamma \left( e_{0},c_{0}\right) =\left\{ \mathbf{u}\in X\ |\ E(%
\mathbf{u})=e_{0},\ \left\vert C(\mathbf{u})\right\vert =c_{0}\right\}
\label{plis}
\end{equation}%
where%
\begin{equation}
e_{0}=\min \left\{ E(\mathbf{u})\ |\ \left\vert C(\mathbf{u})\right\vert
=c_{0}\right\} .  \label{minbis}
\end{equation}
\end{definition}

Notice that, by (\ref{minbis}), we have that a hylomorphic soliton $\mathbf{u%
}_{0}$ satisfies the following nonlinear eigenvalue problem:%
\begin{equation*}
E^{\prime }(\mathbf{u}_{0})=\lambda C^{\prime }(\mathbf{u}_{0}).
\end{equation*}

In general, a minimizer $\mathbf{u}_{0}$ of $E$ on $\mathfrak{M}_{c_{_{0}}}=$
$\left\{ \mathbf{u}\in X\ \ \left\vert C(\mathbf{u})\right\vert
=c_{0}\right\} $ is not a soliton; in fact, according to Def. \ref{ds}, it
is necessary to prove the following facts:

\begin{itemize}
\item (i) The set $\Gamma \left( e_{0},c_{0}\right) $ is stable.

\item (ii) The set $\Gamma \left( e_{0},c_{0}\right) $ is $\mathcal{T}$%
-compact (i.e. compact up to translations).

\item (iii) $0\notin \Gamma \left( e_{0},c_{0}\right) ,$ since otherwise,
some $\mathbf{u}\in \Gamma \left( e_{0},c_{0}\right) $ is not even a
solitary wave (see Def. \ref{solw},(i)).
\end{itemize}

In concrete cases, the point (i) is the most delicate point to prove. If (i)
does not hold, according to our definitions, $\mathbf{u}_{0}$ is a solitary
wave but not a soliton.

Now let us see the general definition of hylomorphic soliton.

\bigskip

\begin{definition}
\label{boh}Let $\left( X,\gamma \right) $ be a dynamical system of type FT
and invariant for the action of a Lie group $G$, namely, for any $\mathbf{u}%
\in X,\ \forall g\in G,$ 
\begin{equation*}
g\gamma _{t}\mathbf{u=}\gamma _{t}g\mathbf{u.}
\end{equation*}%
$\mathbf{u}$ is called \textbf{hylomorphic soliton }if $\mathbf{u=}g\mathbf{u%
}_{0}$ where $\mathbf{u}_{0}$ is a standing hylomorphic soliton and $g$ is a
suitable element of $G.$
\end{definition}

\bigskip

In the application $G$ will be a representation of the Galileo or of the
Lorentz group. Now let us illustrate with an example Def. \ref{tdc} and Def. %
\ref{boh}.

\textbf{Example}. Let us consider the example (\ref{stand})\textbf{. }The
standing wave $u(x)e^{-i\omega t}$ is a hylomorphic soliton if 
\begin{equation*}
\Gamma _{\mathbf{0}}=\left\{ u(x-x_{0})e^{-i\theta }\in H^{1}\left( \mathbb{R%
}^{N},\mathbb{C}\right) :x_{0}\in \mathbb{R}^{N};\ \theta \in \mathbb{R}%
\right\}
\end{equation*}%
satisfies the request in Definition \ref{ds} and if $\Gamma _{\mathbf{0}%
}=\Gamma \left( e_{0},c_{0}\right) $ (see (\ref{plis})) for a suitable $%
c_{0}.$

\section{Existence results of hylomorphic solitons\label{cm}}

In the previous section, we have seen that the existence of hylomorphic
soliton is related to the existence of minimizers of the energy. In this
section we will investigate the following minimization problem%
\begin{equation}
\min_{\mathbf{u}\in \mathfrak{M}_{c}}E(\mathbf{u})\ \ 
\text{where\ \ }\mathfrak{M}_{c}:=\left\{ \mathbf{u}\in X\ |\ \left\vert C(%
\mathbf{u})\right\vert =c\right\}  \label{MinPr}
\end{equation}%
and under which conditions the set of the minimizers 
\begin{equation*}
\Gamma \left( e,c\right) =\left\{ \mathbf{u}\in X\ |\ E(\mathbf{u})=e,\
\left\vert C(\mathbf{u})\right\vert =c\right\} ;\ e=\min_{\mathbf{u}\in 
\mathfrak{M}_{c}}E(\mathbf{u})
\end{equation*}%
is stable.

\subsection{The abstract framework}

The following definitions could be given in a more abstract framework.
Nevertheless, for the sake of definitess, in the following we shall assume
that 
\begin{equation*}
(X,\gamma )\text{ is a dynamical system of FT-type (see Def \ref{ft}})
\end{equation*}%
and that 
\begin{equation*}
G\text{ is a subgroup of }\mathcal{T}\text{ (see(\ref{gg})).}
\end{equation*}

\begin{definition}
A subset $\Gamma \subset X$ is called $G$-invariant if 
\begin{equation*}
\forall \mathbf{u}\in \Gamma ,\ \forall g\in G,\ g\mathbf{u}\in \Gamma .
\end{equation*}
\end{definition}

\begin{definition}
A sequence $\mathbf{u}_{n}$ in $X$ is called $G$\emph{-compact }if there is
a subsequence $\mathbf{u}_{n_{k}}$ and a sequence $g_{k}\in G$ such that $%
g_{k}\mathbf{u}_{n_{k}}$ is convergent. A subset $\Gamma \subset X$ is
called $G$-compact if every sequence in $\Gamma $ is $G$\emph{-compact.}
\end{definition}

Observe that the above definition reduces to Definition \ref{carla} if $G=%
\mathcal{T}$. If $G=\left\{ Id\right\} $ or, more in general, it is a
compact group, $G$\emph{-}compactness implies compactness. If $G$ is not
compact such as the tranlation group $\mathcal{T}$, $G$\emph{-}compactness
is a weaker notion than compactness.

\begin{definition}
\label{gcompattoa}A $G$-invariant functional $J$ on $X$ is called $G$%
-compact if any minimizing sequence $\mathbf{u}_{n}$ is $G$-compact.
\end{definition}

Clearly a $G$-compact functional has a $G$-compact set of minimizers.

\begin{definition}
\label{sp}We say that a functional $F$ on $X$ has the splitting property if
given a sequence $\mathbf{u}_{n}=\mathbf{u}+\mathbf{w}_{n}\in X$ such that $%
\mathbf{w}_{n}$ converges weakly to $0$, we have that 
\begin{equation*}
F(\mathbf{u}_{n})=F(\mathbf{u})+F(\mathbf{w}_{n})+o(1).
\end{equation*}
\end{definition}

\begin{remark}
\label{quadratic}Every quadratic form, which is continuous and symmetric,
satisfies the splitting property; in fact, in this case, we have that $F(%
\mathbf{u}):=\left\langle L\mathbf{u},\mathbf{u}\right\rangle $ for some
continuous selfajoint operator $L;$ then, given a sequence $\mathbf{u}_{n}=%
\mathbf{u}+\mathbf{w}_{n}$ with $\mathbf{w}_{n}\rightharpoonup 0$ weakly, we
have that%
\begin{align*}
F(\mathbf{u}_{n})& =\left\langle L\mathbf{u},\mathbf{u}\right\rangle
+\left\langle L\mathbf{w}_{n},\mathbf{w}_{n}\right\rangle +2\left\langle L%
\mathbf{u},\mathbf{w}_{n}\right\rangle \\
& =F(\mathbf{u})+F(\mathbf{w}_{n})+o(1).
\end{align*}
\end{remark}

\begin{definition}
\label{na}\textbf{\ }A sequence $\mathbf{u}_{n}\in X$ is called G-\textbf{%
vanishing sequence}\textit{\ if it is bounded and if for any subsequence }$%
\mathbf{u}_{n_{k}}$ and for any sequence $g_{_{k}}\subset G$ the sequence $%
g_{k}\mathbf{u}_{n_{k}}$ converges weakly to $0.$
\end{definition}

So, if $\mathbf{u}_{n}\rightarrow 0$ strongly, $\mathbf{u}_{n}$ is a
G-vanishing sequence. However, if $\mathbf{u}_{n}\rightharpoonup 0$ weakly,
it might happen that it is not a G-vanishing sequence; namely it might exist
a subsequence $\mathbf{u}_{n_{k}}$ and a sequence $g_{k}\subset G$ such that 
$g_{k}\mathbf{u}_{n_{k}}$ is weakly convergent to some $\mathbf{\bar{u}}\neq
0$. Let see an example; if $\mathbf{u}_{0}\in X$ is a solitary wave and $%
t_{n}\rightarrow +\infty ,$ then the sequence $\gamma _{t_{n}}\mathbf{u}_{0}$
is not a $\mathcal{T}$-vanishing sequence.

In the following $E$ and $C$ will denote two constants of the motion for the
dynamical system (in the applications they will be the energy and the
charge). We will assume that 
\begin{equation*}
E\text{ and }C\text{ are }C^{1}\text{ and bounded functionals on }X.
\end{equation*}

We set 
\begin{equation}
\Lambda \left( \mathbf{u}\right) :=\frac{E\left( \mathbf{u}\right) }{%
\left\vert C\left( \mathbf{u}\right) \right\vert },  \label{lambda}
\end{equation}

Since $E$ and $C$ are constants of motion, also $\Lambda $ is a constant of
motion; it will be called \textbf{hylenic ratio} (see the definition of
charge, sec. \ref{cl}) and, as we will see it will play a central role in
this theory.

The notions of G-vanishing sequence and of hylenic ratio allow to introduce
the following (important) definition:

\begin{definition}
\label{dhc}We say that the \textit{hylomorphy condition holds%
\index{hylomorphy condition} }if%
\begin{equation}
\underset{\mathbf{u}\in X}{\inf }%
\frac{E\left( \mathbf{u}\right) }{\left\vert C\left( \mathbf{u}\right)
\right\vert }<\Lambda _{0}.  \label{hh}
\end{equation}%
where%
\begin{equation}
\Lambda _{0}:=\ \inf \left\{ \lim \inf \ \Lambda (\mathbf{u}_{n})\ |\ 
\mathbf{u}_{n}\ \text{is a G-vanishing sequence}\right\} .  \label{hylo}
\end{equation}%
Moreover, we say that $\mathbf{u}_{0}\in X$ satisfies the \textit{hylomorphy
condition }if, 
\begin{equation}
\frac{E\left( \mathbf{u}_{0}\right) }{\left\vert C\left( \mathbf{u}%
_{0}\right) \right\vert }<\Lambda _{0}.  \label{yc}
\end{equation}
\end{definition}

By this definition, using the above notation, we have the following:%
\begin{equation*}
\lim \Lambda \left( \mathbf{u}_{n}\right) <\Lambda _{0}\Rightarrow \exists 
\mathbf{u}_{n_{k}},g_{k}\in G:g_{k}\mathbf{u}_{n_{k}}\rightharpoonup \mathbf{%
\bar{u}}\neq 0.
\end{equation*}

In order to apply the existence theorems of sect. \ref{saet}, it is
necessary to estimate $\Lambda _{0};$ the following propositons may help to
do this.

\begin{proposition}
\label{diesis}Assume that there exists a seminorm $\left\Vert {}\right\Vert
_{\sharp }$ on $X$ such that\label{kk} 
\begin{equation}
\left\{ \mathbf{u}_{n}\ \text{is a }G-\text{vanishing sequence}\right\}
\Rightarrow \left\Vert \mathbf{u}_{n}\right\Vert _{\sharp }\rightarrow 0.
\label{seminorm}
\end{equation}%
Then%
\begin{equation}
\underset{\left\Vert \mathbf{u}\right\Vert _{\sharp }\rightarrow 0}{\lim
\inf }\ \Lambda (\mathbf{u})\leq \Lambda _{0}\leq \ \underset{\left\Vert 
\mathbf{u}\right\Vert \rightarrow 0}{\lim \inf }\ \Lambda (\mathbf{u}).
\label{aaaa}
\end{equation}
\end{proposition}

\textbf{Proof. }It follows directly from the definition (\ref{hylo}) of $%
\Lambda _{0}$ and (\ref{seminorm}).

\bigskip $\square $

\begin{proposition}
If $E$ and $C$ are twice differentiable in $0$ and 
\begin{equation*}
E(0)=C(0)=0;\ E^{\prime }(0)=C^{\prime }(0)=0,
\end{equation*}%
then we have that%
\begin{equation*}
\Lambda _{0}\leq \ \underset{\mathbf{u}\neq 0}{\inf }\frac{E^{\prime \prime
}(0)\left[ \mathbf{u,u}\right] }{\left\vert C^{\prime \prime }(0)\left[ 
\mathbf{u,u}\right] \right\vert }.
\end{equation*}
\end{proposition}

\textbf{Proof.} By the above proposition,\textbf{\ }%
\begin{eqnarray*}
\Lambda _{0} &\leq &\ \underset{\left\Vert \mathbf{u}\right\Vert \rightarrow
0}{\lim \inf }\ \Lambda (\mathbf{u})=\ \underset{\left\Vert \mathbf{u}%
\right\Vert \rightarrow 0}{\lim \inf }\frac{E(0)+E^{\prime }(0)\left[ 
\mathbf{u}\right] +E^{\prime \prime }(0)\left[ \mathbf{u,u}\right]
+o(\left\Vert \mathbf{u}\right\Vert ^{2})}{\left\vert C(0)+C^{\prime }(0)%
\left[ \mathbf{u}\right] +C^{\prime \prime }(0)\left[ \mathbf{u,u}\right]
+o(\left\Vert \mathbf{u}\right\Vert ^{2})\right\vert } \\
&=&\ \underset{\mathbf{u}\neq 0}{\inf }\frac{E^{\prime \prime }(0)\left[ 
\mathbf{u,u}\right] }{\left\vert C^{\prime \prime }(0)\left[ \mathbf{u,u}%
\right] \right\vert }.
\end{eqnarray*}

$\square $

\bigskip

Now, finally, we can give some abstract theorems relative to the existence
of hylomorphic solitons.

\subsection{Statement of the abstract existence theorems\label{saet}}

We formulate the assumptions on $E$ and $C$:

\begin{itemize}
\item \textit{(EC-0) \textbf{(Values at 0)}}%
\begin{equation*}
E(0)=C(0)=0;\ E^{\prime }(0)=C^{\prime }(0)=0.\ 
\end{equation*}

\item \textit{(EC-1)\textbf{(Invariance)} }$E(\mathbf{u})$\textit{\ and }$C(%
\mathbf{u})$\textit{\ are }$G$\textit{-invariant.}

\item \textit{(EC-2)\textbf{(Splitting property)} }$E$\textit{\ and }$C$%
\textit{\ satisfy the splitting property.}

\item \textit{(EC-3)\textbf{(Coercivity) }We assume that}

\begin{itemize}
\item (i) \textit{\textbf{\ }}$\forall \mathbf{u}\neq 0,\ E(\mathbf{u})>0.$

\item (ii) \textit{if }$\left\Vert \mathbf{u}_{n}\right\Vert \rightarrow
\infty ,\ $\textit{then} $E(\mathbf{u}_{n})\rightarrow \infty ;$

\item (iii) \textit{if }$E(\mathbf{u}_{n})\rightarrow 0$, \textit{then} $%
\left\Vert \mathbf{u}_{n}\right\Vert \rightarrow 0.$
\end{itemize}
\end{itemize}

Now we can state the main results:

\begin{theorem}
\label{astra1}Assume that $E\ $and $C$ satisfy (EC-0),...,(EC-2) and \textit{%
(EC-3). Moreover assume that }the hylomorphy condition of Def. \ref{dhc} is
satisfied. Then there exists a family of hylomorphic solitons $\mathbf{u}%
_{\delta },$ $\delta \in \left( 0,\delta _{\infty }\right) ,$ $\delta
_{\infty }>0$.
\end{theorem}

\begin{theorem}
\label{astra1+}Let the assumptions of theorem \ref{astra1} hold. Moreover
assume that 
\begin{equation}
\left\Vert E^{\prime }(\mathbf{u})\right\Vert +\left\Vert C^{\prime }(%
\mathbf{u})\right\Vert =0\Leftrightarrow \mathbf{u}=0.  \label{poo}
\end{equation}

Then for every $\delta \in \left( 0,\delta _{\infty }\right) ,$ $\delta
_{\infty }>0,\ $there exists a hylomorphic soliton $\mathbf{u}_{\delta }$.
Moreover, if $\delta _{1}<\delta _{2},$ the corresponding solitons $\mathbf{u%
}_{\delta _{1}},\mathbf{u}_{\delta _{2}}$ are distinct, namely we have that

\begin{itemize}
\item (a) $\Lambda (\mathbf{u}_{\delta _{1}})<\Lambda (\mathbf{u}_{\delta
_{2}})$

\item (b) $\left\vert C(\mathbf{u}_{\delta _{1}})\right\vert >\left\vert C(%
\mathbf{u}_{\delta _{2}})\right\vert .$

\item (c) $E(\mathbf{u}_{\delta _{1}})>E(\mathbf{u}_{\delta _{2}})$
\end{itemize}
\end{theorem}

The proofs of the above results are in the remaining part of this section.
In subsection \ref{33} we prove the existence of minimizers, namely that $%
\Gamma (e,c)\neq \varnothing $ (see (\ref{plis})) and in subsection \ref{35}%
, we prove the stability of $\Gamma (e,c),$ namely that the minimizers are
hylomorphic solitons.

\subsection{A minimization result\label{33}}

We start with a technical lemma.

\begin{lemma}
\label{leman}Let $\mathbf{u}_{n}=\mathbf{u}+\mathbf{w}_{n}\in X$ be a
sequence such that $\mathbf{w}_{n}$ converges weakly to $0$. Then, up to a
subsequence, we have 
\begin{equation*}
\lim \Lambda \left( \mathbf{u+\mathbf{w}_{n}}\right) \geq \min \left(
\Lambda \left( \mathbf{u}\right) ,\lim \Lambda \left( \mathbf{\mathbf{w}_{n}}%
\right) \right)
\end{equation*}%
and the equality holds if and only if $\Lambda \left( \mathbf{u}\right)
=\lim \Lambda \left( \mathbf{\mathbf{w}_{n}}\right) .$
\end{lemma}

\textbf{Proof.} Given four real numbers $A,B,a,b,$ (with $B,b>0$), we have
that%
\begin{equation}
\frac{A+a}{B+b}\geq \min \left( \frac{A}{B},\frac{a}{b}\right) .
\label{figa}
\end{equation}%
In fact, suppose that $\frac{A}{B}\geq \frac{a}{b};$ then%
\begin{equation*}
\frac{A+a}{B+b}=\frac{\frac{A}{B}B+\frac{a}{b}b}{B+b}\geq \frac{\frac{a}{b}B+%
\frac{a}{b}b}{B+b}=\frac{a}{b}\geq \min \left( \frac{A}{B},\frac{a}{b}%
\right) .
\end{equation*}%
Notice that the equality holds if and only if $\frac{A}{B}=\frac{a}{b}.$
Using the splitting property and the above inequality, up to a subsequence,
we have that%
\begin{eqnarray*}
\lim \Lambda \left( \mathbf{u+\mathbf{w}_{n}}\right) &=&\frac{\lim E\left( 
\mathbf{u+\mathbf{w}_{n}}\right) }{\lim \left\vert C\left( \mathbf{u+\mathbf{%
w}_{n}}\right) \right\vert }=\frac{E\left( \mathbf{u}\right) +\lim E\left( 
\mathbf{\mathbf{w}_{n}}\right) }{\left\vert C\left( \mathbf{u}\right) +\lim
C\left( \mathbf{\mathbf{w}_{n}}\right) \right\vert }\geq \\
&&\frac{E\left( \mathbf{u}\right) +\lim E\left( \mathbf{\mathbf{w}_{n}}%
\right) }{\left\vert C\left( \mathbf{u}\right) \right\vert +\left\vert \lim
C\left( \mathbf{\mathbf{w}_{n}}\right) \right\vert } \\
&\geq &\min \left( \frac{E\left( \mathbf{u}\right) }{\left\vert C\left( 
\mathbf{u}\right) \right\vert },\frac{\lim E\left( \mathbf{\mathbf{w}_{n}}%
\right) }{\lim \left\vert C\left( \mathbf{\mathbf{w}_{n}}\right) \right\vert 
}\right) =\min \left( \Lambda \left( \mathbf{u}\right) ,\lim \Lambda \left( 
\mathbf{\mathbf{w}_{n}}\right) \right) .
\end{eqnarray*}

$\square $

For any $\delta >0,$ set%
\begin{equation*}
J_{\delta }(\mathbf{u})=\Lambda \left( \mathbf{u}\right) +\delta E(\mathbf{u}%
)
\end{equation*}%
By the hylomorphy condition (\ref{hh}) we have%
\begin{equation}
\delta _{\infty }=\sup \left\{ \delta >0\ |\ \exists \mathbf{v}:\Lambda
\left( \mathbf{v}\right) +\delta E(\mathbf{v})<\Lambda _{0}\ \right\} \in 
\mathbb{R}^{+}\cup \left\{ +\infty \right\} .  \label{tino}
\end{equation}

Clearly, if $\delta \in \left[ 0,\delta _{\infty }\right) ,$ $\exists 
\mathbf{v}:\Lambda \left( \mathbf{v}\right) +\delta E(\mathbf{v})<\Lambda
_{0}.$

\begin{theorem}
\label{legicomp}Assume that $E\ $and $C$ satisfy (EC-0),...,(EC-3) and the
hylomorphy condition (\ref{hh}). Then, for every $\delta \in \left( 0,\delta
_{\infty }\right) ,$ $J_{\delta }$ is $G$-compact and it has a minimizer $%
\mathbf{u}_{\delta }\neq 0.$ Moreover, $\mathbf{u}_{\delta }\in \Gamma
\left( e_{\delta },c_{\delta }\right) $ (see (\ref{plis})) where $e_{\delta
}=E(\mathbf{u}_{\delta }),$ $c_{\delta }=\left\vert C(\mathbf{u}_{\delta
})\right\vert >0$.
\end{theorem}

\textbf{Proof.} Let $\mathbf{u}_{n}$ be a minimizing sequence of $J_{\delta
} $ ($\delta \in (0,\delta _{\infty })).$ This sequence $\mathbf{u}_{n}$ is
bounded in $X.$ In fact, arguing by contradiction, assume that, up to a
subsequence, $\left\Vert \mathbf{u}_{n}\right\Vert $ $\rightarrow \infty .$
Then by (EC-3) (ii), $E(\mathbf{u}_{n})\rightarrow \infty $ and hence $%
J_{\delta }(\mathbf{u}_{n})\rightarrow \infty $ which contradicts the fact
that $\mathbf{u}_{n}$ is a minimizing sequence of $J_{\delta }.$

We now set%
\begin{equation*}
j_{\delta }:=\ \underset{\mathbf{u}\in X}{\inf }J_{\delta }\left( \mathbf{u}%
\right) .
\end{equation*}%
Since $\delta \in (0,\delta _{\infty })$, where $\delta _{\infty }$ is
defined in (\ref{tino}), we have that 
\begin{equation}
j_{\delta }<\Lambda _{0}.  \label{favino}
\end{equation}%
Moreover, since $E\geq 0,$we have%
\begin{equation*}
0\leq \Lambda \left( \mathbf{u}_{n}\right) \leq J_{\delta }(\mathbf{u}_{n})
\end{equation*}%
and%
\begin{equation*}
J_{\delta }(\mathbf{u}_{n})\rightarrow j_{\delta }<\Lambda _{0}.
\end{equation*}%
Then, up to a subsequence$,$ $\Lambda \left( \mathbf{u}_{n}\right)
\rightarrow \lambda <\Lambda _{0}.$ So, by definition (\ref{hylo}) of $%
\Lambda _{0\text{ }}$, $\mathbf{u}_{n}$ is not a $G-$vanishing sequence.
Hence, by Def. \ref{na}, we can extract a subsequence $\mathbf{u}_{n_{k}}$
and we can take a sequence $g_{k}\subset G$ such that $\mathbf{u}%
_{k}^{\prime }:=g_{k}\mathbf{u}_{n_{k}}$ is weakly convergent to some 
\begin{equation}
\mathbf{u}_{\delta }\neq 0.  \label{fi}
\end{equation}%
We can write 
\begin{equation*}
\mathbf{u}_{n}^{\prime }=\mathbf{u}_{\delta }+\mathbf{w}_{n}
\end{equation*}%
with $\mathbf{w}_{n}\rightharpoonup 0$ weakly. We want to prove that $%
\mathbf{w}_{n}\rightarrow 0$ strongly.

By lemma \ref{leman} and by the splitting property of $E$, we have, up to a
subsequence, that%
\begin{eqnarray*}
j_{\delta } &=&\lim J_{\delta }\left( \mathbf{u}_{\delta }+\mathbf{w}%
_{n}\right) =\lim \left[ \Lambda \left( \mathbf{u}_{\delta }+\mathbf{w}%
_{n}\right) +\delta E\left( \mathbf{u}_{\delta }+\mathbf{w}_{n}\right) %
\right] \\
&\geq &\left[ \min \left\{ \Lambda \left( \mathbf{u}_{\delta }\right) ,\lim
\Lambda \left( \mathbf{w}_{n}\right) \right\} \right] +\delta E\left( 
\mathbf{u}_{\delta }\right) +\delta \lim E\left( \mathbf{w}_{n}\right) .
\end{eqnarray*}

Now there are two possibilities (up to subsequences): 1- $\min \left\{
\Lambda \left( \mathbf{u}_{\delta }\right) ,\lim \Lambda \left( \mathbf{w}%
_{n}\right) \right\} =\lim \Lambda \left( \mathbf{w}_{n}\right) $;\ 2- $\
\min \left\{ \Lambda \left( \mathbf{u}_{\delta }\right) ,\lim \Lambda \left( 
\mathbf{w}_{n}\right) \right\} =\Lambda \left( \mathbf{u}_{\delta }\right) .$
We will show that the possibility 1 cannot occur. In fact, if it holds, we
have that%
\begin{eqnarray*}
j_{\delta } &\geq &\lim \Lambda \left( \mathbf{w}_{n}\right) +\delta E\left( 
\mathbf{u}_{\delta }\right) +\delta \lim E\left( \mathbf{w}_{n}\right) \\
&=&\lim J_{\delta }\left( \mathbf{w}_{n}\right) +\delta E\left( \mathbf{u}%
_{\delta }\right) \\
&\geq &j_{\delta }+\delta E\left( \mathbf{u}_{\delta }\right)
\end{eqnarray*}%
and hence, we get that $E\left( \mathbf{u}_{\delta }\right) \leq 0,$
contradicting (\ref{fi}). Then possibility 2 holds and we have that%
\begin{eqnarray*}
j_{\delta } &\geq &\Lambda \left( \mathbf{u}_{\delta }\right) +\delta
E\left( \mathbf{u}_{\delta }\right) +\delta \lim E\left( \mathbf{w}%
_{n}\right) \\
&=&J_{\delta }\left( \mathbf{u}_{\delta }\right) +\delta \lim E\left( 
\mathbf{w}_{n}\right) \\
&\geq &j_{\delta }+\delta \lim E\left( \mathbf{w}_{n}\right) .
\end{eqnarray*}%
Then, $\lim E\left( \mathbf{w}_{n}\right) =0$ and, by (EC-3)(iii), $\mathbf{w%
}_{n}\rightarrow 0$ strongly. Then $J_{\delta }\left( \mathbf{u}_{n}^{\prime
}\right) \rightarrow J_{\delta }\left( \mathbf{u}_{\delta }\right) .$ So $%
J_{\delta }$ is $G.$-compact and $\mathbf{u}_{\delta }$ is a minimizer.

To prove the second part of the theorem, we set:.

\begin{eqnarray*}
e_{\delta } &=&E(\mathbf{u}_{\delta }) \\
c_{\delta } &=&\left\vert C(\mathbf{u}_{\delta })\right\vert \\
\mathfrak{M}_{\delta } &:&=\left\{ \mathbf{u}\in X\ |\ \left\vert C(\mathbf{u%
})\right\vert =c_{\delta }\right\} .
\end{eqnarray*}%
Since 
\begin{equation*}
J_{\delta }|_{\mathfrak{M}_{\delta }}=\frac{E}{c_{\delta }}+\delta E=\left( 
\frac{1}{c_{\delta }}+\delta \right) E
\end{equation*}%
it follows that $\mathbf{u}_{\delta }$ minimizes also $E|_{\mathfrak{M}%
_{\delta }}$.

$\square $

In the following $\mathbf{u}_{\delta }$ will denote a minimizer of $%
J_{\delta }.$

\begin{lemma}
\label{MM}Let the assumptions of Theorem \ref{legicomp} be satisfied. If $%
\delta _{1},\delta _{2}\in (0,\delta _{\infty })$ $\delta _{1}<\delta _{2}$ (%
$\delta _{\infty }$ as in (\ref{tino})), then the minimizers $\mathbf{u}%
_{\delta _{1}},\mathbf{u}_{\delta _{2}}$ of $J_{\delta _{1}},$ $J_{\delta
_{2}}$ respectively satisfy the following inequalities:

\begin{itemize}
\item (a) $J_{\delta _{1}}(\mathbf{u}_{\delta _{1}})<J_{\delta _{2}}(\mathbf{%
u}_{\delta _{2}})$

\item (b) $E(\mathbf{u}_{\delta _{1}})\geq E(\mathbf{u}_{\delta _{2}}),\ $

\item (c) $\Lambda (\mathbf{u}_{\delta _{1}})\leq \Lambda (\mathbf{u}%
_{\delta _{2}}),$

\item (d) $\left\vert C(\mathbf{u}_{\delta _{1}})\right\vert \geq \left\vert
C(\mathbf{u}_{\delta _{2}})\right\vert .$
\end{itemize}
\end{lemma}

\textbf{Proof.} (a) 
\begin{eqnarray*}
J_{\delta _{1}}\left( \mathbf{u}_{\delta _{1}}\right) &=&\Lambda \left( 
\mathbf{u}_{\delta _{1}}\right) +\delta _{1}E(\mathbf{u}_{\delta _{1}}) \\
&\leq &\Lambda \left( \mathbf{u}_{\delta _{2}}\right) +\delta _{1}E(\mathbf{u%
}_{\delta _{2}})\ \ (\text{since }\mathbf{u}_{\delta _{1}}\ \text{minimizes }%
J_{\delta _{1}}\text{)} \\
&<&\Lambda \left( \mathbf{u}_{\delta _{2}}\right) +\delta _{2}E(\mathbf{u}%
_{\delta _{2}})\ \ (\text{since }E\ \text{is positive)} \\
&=&J_{\delta _{2}}\left( \mathbf{u}_{\delta _{2}}\right) .
\end{eqnarray*}%
(b) We set 
\begin{eqnarray*}
\Lambda (\mathbf{u}_{\delta _{1}}) &=&\Lambda (\mathbf{u}_{\delta _{2}})+a \\
E(\mathbf{u}_{\delta _{1}}) &=&E(\mathbf{u}_{\delta _{2}})+b.
\end{eqnarray*}%
We need to prove that $b\geq 0\ $and $a\leq 0.$ We have%
\begin{eqnarray}
J_{\delta _{2}}\left( \mathbf{u}_{\delta _{2}}\right) &\leq &J_{\delta _{2}}(%
\mathbf{u}_{\delta _{1}})\Rightarrow  \notag \\
\Lambda \left( \mathbf{u}_{\delta _{2}}\right) +\delta _{2}E(\mathbf{u}%
_{\delta _{2}}) &\leq &\Lambda \left( \mathbf{u}_{\delta _{1}}\right)
+\delta _{2}E(\mathbf{u}_{\delta _{1}})\Rightarrow  \notag \\
\Lambda \left( \mathbf{u}_{\delta _{2}}\right) +\delta _{2}E(\mathbf{u}%
_{\delta _{2}}) &\leq &\left( \Lambda \left( \mathbf{u}_{\delta _{2}}\right)
+a\right) +\delta _{2}\left( E(\mathbf{u}_{\delta _{2}})+b\right) \Rightarrow
\notag \\
0 &\leq &a+\delta _{2}b.  \label{blu}
\end{eqnarray}

On the other hand,%
\begin{eqnarray}
J_{\delta _{1}}\left( \mathbf{u}_{\delta _{2}}\right) &\geq &J_{\delta _{1}}(%
\mathbf{u}_{\delta _{1}})\Rightarrow  \notag \\
\Lambda \left( \mathbf{u}_{\delta _{2}}\right) +\delta _{1}E(\mathbf{u}%
_{\delta _{2}}) &\geq &\Lambda \left( \mathbf{u}_{\delta _{1}}\right)
+\delta _{1}E(\mathbf{u}_{\delta _{1}})\Rightarrow  \notag \\
\Lambda \left( \mathbf{u}_{\delta _{2}}\right) +\delta _{1}E(\mathbf{u}%
_{\delta _{2}}) &\geq &\left( \Lambda \left( \mathbf{u}_{\delta _{2}}\right)
+a\right) +\delta _{1}\left( E(\mathbf{u}_{\delta _{2}})+b\right) \Rightarrow
\notag \\
0 &\geq &a+\delta _{1}b.  \label{verde}
\end{eqnarray}

From (\ref{blu}) and (\ref{verde}) we get%
\begin{equation*}
\left( \delta _{2}-\delta _{1}\right) b\geq 0
\end{equation*}%
and hence $b\geq 0.$

Moreover by (\ref{blu}) and (\ref{verde}) we also get%
\begin{equation*}
\left( \frac{1}{\delta _{2}}-\frac{1}{\delta _{1}}\right) a\geq 0
\end{equation*}%
and hence $a\leq 0.$ Since $\left\vert C(\mathbf{u})\right\vert =\frac{E(%
\mathbf{u})}{\Lambda (\mathbf{u})},\ $also inequality (d) follows.

$\square $

\begin{lemma}
\label{BB} Let the assumptions of Theorem \ref{legicomp} be satisfied and
assume that also (\ref{poo}) is satisfied. If $\delta _{1},\delta _{2}\in
(0,\delta _{\infty })$ ($\delta _{\infty }$ as in (\ref{tino})), $\delta
_{1}<\delta _{2}$, then the minimizers $\mathbf{u}_{\delta _{1}},\mathbf{u}%
_{\delta _{2}}$ of $J_{\delta _{1}},$ $J_{\delta _{2}}$ respectively satisfy
the following inequalities:

\begin{itemize}
\item (a) $E(\mathbf{u}_{\delta _{1}})>E(\mathbf{u}_{\delta _{2}}),\ $

\item (b) $\Lambda (\mathbf{u}_{\delta _{1}})<\Lambda (\mathbf{u}_{\delta
_{2}})\ $

\item (c) $\left\vert C(\mathbf{u}_{\delta _{1}})\right\vert >\left\vert C(%
\mathbf{u}_{\delta _{2}})\right\vert .$
\end{itemize}
\end{lemma}

\textbf{Proof:} Let $\delta _{1},\delta _{2}\in \left( 0,\delta _{\infty
}\right) $ and assume that $\delta _{1}<\delta _{2}.$

(a) It is sufficient to prove that $E(\mathbf{u}_{\delta _{1}})\neq E(%
\mathbf{u}_{\delta _{2}}).$ We argue indirectly and assume that 
\begin{equation}
E(\mathbf{u}_{\delta _{1}})=E(\mathbf{u}_{\delta _{2}}).  \label{bla}
\end{equation}

By the previous lemma, we have that%
\begin{equation}
\Lambda \left( \mathbf{u}_{\delta _{1}}\right) \leq \Lambda \left( \mathbf{u}%
_{\delta _{2}}\right) .  \label{ble}
\end{equation}%
Also, we have that%
\begin{eqnarray*}
\Lambda \left( \mathbf{u}_{\delta _{2}}\right) +\delta _{2}E\left( \mathbf{u}%
_{\delta _{2}}\right) &\leq &\Lambda \left( \mathbf{u}_{\delta _{1}}\right)
+\delta _{2}E(\mathbf{u}_{\delta _{1}})\ \ \text{(since\ }\mathbf{u}_{\delta
_{2}}\ \text{minimizes }J_{\delta _{2}}\text{)} \\
&=&\Lambda \left( \mathbf{u}_{\delta _{1}}\right) +\delta _{2}E\left( 
\mathbf{u}_{\delta _{2}}\right) \ \ \text{(by\ (\ref{bla}))}
\end{eqnarray*}%
and so%
\begin{equation*}
\Lambda \left( \mathbf{u}_{\delta _{2}}\right) \leq \Lambda \left( \mathbf{u}%
_{\delta _{1}}\right)
\end{equation*}%
and by (\ref{ble}) we get 
\begin{equation}
\Lambda \left( \mathbf{u}_{\delta _{1}}\right) =\Lambda \left( \mathbf{u}%
_{\delta _{2}}\right) .  \label{bli}
\end{equation}

Then, it follows that $\mathbf{u}_{\delta _{1}}$ is a minimizer of $%
J_{\delta _{2}};$in fact, by (\ref{bli}) and (\ref{bla}))%
\begin{eqnarray*}
J_{\delta _{2}}\left( \mathbf{u}_{\delta _{1}}\right) &=&\Lambda \left( 
\mathbf{u}_{\delta _{1}}\right) +\delta _{2}E\left( \mathbf{u}_{\delta
_{1}}\right) \\
&=&\Lambda \left( \mathbf{u}_{\delta _{2}}\right) +\delta _{2}E\left( 
\mathbf{u}_{\delta _{2}}\right) =J_{\delta _{2}}\left( \mathbf{u}_{\delta
_{2}}\right) .
\end{eqnarray*}%
Then, we have that $J_{\delta _{2}}^{\prime }\left( \mathbf{u}_{\delta
_{1}}\right) =0$ as well as $J_{\delta _{1}}^{\prime }\left( \mathbf{u}%
_{\delta _{1}}\right) =0$ which esplicitely give%
\begin{eqnarray*}
\Lambda ^{\prime }\left( \mathbf{u}_{\delta _{1}}\right) +\delta
_{2}E^{\prime }\left( \mathbf{u}_{\delta _{1}}\right) &=&0 \\
\Lambda ^{\prime }\left( \mathbf{u}_{\delta _{1}}\right) +\delta
_{1}E^{\prime }\left( \mathbf{u}_{\delta _{1}}\right) &=&0.
\end{eqnarray*}%
The above equations imply that $E^{\prime }\left( \mathbf{u}_{\delta
_{1}}\right) =0$ and $\Lambda ^{\prime }\left( \mathbf{u}_{\delta
_{1}}\right) =0,$ and since $\Lambda \left( \mathbf{u}\right) =\frac{E\left( 
\mathbf{u}\right) }{\left\vert C\left( \mathbf{u}\right) \right\vert },$ we
get that $C^{\prime }\left( \mathbf{u}_{\delta _{1}}\right) =0.$ Then by (%
\ref{poo}) $\mathbf{u}_{\delta _{1}}=0,$ and this fact contradicts Th. \ref%
{legicomp}.

(b) We argue indirectly and assume that 
\begin{equation}
\Lambda (\mathbf{u}_{\delta _{1}})=\Lambda (\mathbf{u}_{\delta _{2}}).
\label{bla2}
\end{equation}

By (a), we have that%
\begin{equation}
E\left( \mathbf{u}_{\delta _{1}}\right) >E\left( \mathbf{u}_{\delta
_{2}}\right) .  \label{ble2}
\end{equation}%
Also, we have that%
\begin{eqnarray*}
\Lambda \left( \mathbf{u}_{\delta _{1}}\right) +\delta _{1}E\left( \mathbf{u}%
_{\delta _{1}}\right) &\leq &\Lambda \left( \mathbf{u}_{\delta _{2}}\right)
+\delta _{1}E(\mathbf{u}_{\delta _{2}})\ \ \text{(since\ }\mathbf{u}_{\delta
_{1}}\ \text{minimizes }J_{\delta _{1}}\text{)} \\
&=&\Lambda \left( \mathbf{u}_{\delta _{1}}\right) +\delta _{1}E\left( 
\mathbf{u}_{\delta _{2}}\right) \ \ \text{(by\ (\ref{bla2}))}
\end{eqnarray*}%
and so%
\begin{equation*}
E\left( \mathbf{u}_{\delta _{1}}\right) \leq E\left( \mathbf{u}_{\delta
_{2}}\right)
\end{equation*}%
and by (\ref{ble2}) we get a contradiction.

(c) Since%
\begin{equation*}
\left\vert C\left( \mathbf{u}_{\delta }\right) \right\vert =\frac{E\left( 
\mathbf{u}_{\delta }\right) }{\Lambda \left( \mathbf{u}_{\delta }\right) },
\end{equation*}%
the conclusion follows from (a) and (b).

$\square $

\subsection{The stability result\label{35}}

In order to prove Theorem \ref{astra1} it is sufficient to show that the
minimizers in Th. \ref{legicomp} provide solitons, so we have to prove that
the set $\Gamma \left( e,c\right) $ is stable. To do this, we need the (well
known) Liapunov theorem in following form:

\begin{theorem}
\label{propV}Let $\Gamma $ be an invariant set and assume that there exists
a differentiable function $V$ (called Liapunov function) defined on a
neighborhood of $\Gamma $ such that

\begin{itemize}
\item (a) $V(\mathbf{u})\geq0$ and\ $V(\mathbf{u})=0\Leftrightarrow u\in
\Gamma$

\item (b) $\partial_{t}V(\gamma_{t}\left( \mathbf{u}\right) )\leq0$

\item (c) $V(\mathbf{u}_{n})\rightarrow 0\Leftrightarrow d(\mathbf{u}%
_{n},\Gamma )\rightarrow 0.$
\end{itemize}

\noindent Then $\Gamma$ is stable.
\end{theorem}

\textbf{Proof. }For completeness, we give a proof of this well known result.
Arguing by contradiction, assume that $\Gamma,$ satisfying the assumptions
of Th. \ref{propV}, is not stable. Then there exists $\varepsilon>0$ and
sequences $\mathbf{u}_{n}\in X$ and $t_{n}>0$ such that 
\begin{equation}
d(\mathbf{u}_{n},\Gamma)\rightarrow0\text{ and }d(\gamma_{t_{n}}\left( 
\mathbf{u}_{n}\right) ,\Gamma)>\varepsilon.  \label{bingo}
\end{equation}
Then we have%
\begin{equation*}
d(\mathbf{u}_{n},\Gamma)\rightarrow0\Longrightarrow V(\mathbf{u}%
_{n})\rightarrow0\Longrightarrow V(\gamma_{t_{n}}\left( \mathbf{u}%
_{n}\right) )\rightarrow0\Longrightarrow d(\gamma_{t_{n}}\left( \mathbf{u}%
_{n}\right) ,\Gamma)\rightarrow0
\end{equation*}
where the first and the third implications are consequence of property (c).
The second implication follows from property (b). Clearly, this fact
contradicts (\ref{bingo}).

$\square $

\begin{lemma}
\label{LC}Let $V$ be $G$-compact, continuous functional, $V\geq 0$ and let $%
\Gamma =V^{-1}(0)$ be the set of minimizers of $V.\ $If $\Gamma \neq
\varnothing ,$ then $\Gamma $ is $G$-compact and $V$ satisfies the point (c)
of the previous theorem.
\end{lemma}

\textbf{Proof}: The fact that $\Gamma $ is G-compact, is a trivial
consequence of the fact that $\Gamma $ is the set of minimizers of a $G$%
-compact functional $V$. Now we prove (c). First we show the implication $%
\Rightarrow .$ Let $\mathbf{u}_{n}$ be a sequence such that $V(\mathbf{u}%
_{n})\rightarrow 0.$ By contradiction, we assume that $d(\mathbf{u}%
_{n},\Gamma )\nrightarrow 0,$ namely that there is a subsequence $\mathbf{u}%
_{n}^{^{\prime }}$ such that 
\begin{equation}
d(\mathbf{u}_{n}^{\prime },\Gamma )\geq a>0.  \label{kaka}
\end{equation}%
Since $V(\mathbf{u}_{n})\rightarrow 0$ also $V(\mathbf{u}_{n}^{\prime
})\rightarrow 0,$ and, since $V$ is $G$ compact, there exists a sequence $%
g_{n}$ in $G$ such that, for a subsequence $\mathbf{u}_{n}^{\prime \prime }$%
, we have $g_{n}\mathbf{u}_{n}^{\prime \prime }\rightarrow \mathbf{u}_{0}.$
Then 
\begin{equation*}
d(\mathbf{u}_{n}^{\prime \prime },\Gamma )=d(g_{n}\mathbf{u}_{n}^{\prime
\prime },\Gamma )\leq d(g_{n}\mathbf{u}_{n}^{\prime \prime },\mathbf{u}%
_{0})\rightarrow 0
\end{equation*}%
and this contradicts (\ref{kaka}).

Now we prove the other implication $\Leftarrow .$ Let $\mathbf{u}_{n}$ be a
sequence such that $d(\mathbf{u}_{n},\Gamma )\rightarrow 0,$ then there
exists $\mathbf{v}_{n}\in \Gamma $ s.t. 
\begin{equation}
d(\mathbf{u}_{n},\Gamma )\geq d(\mathbf{u}_{n},\mathbf{v}_{n})-\frac{1}{n}.
\label{triplo}
\end{equation}

Since $V$ is G-compact, also $\Gamma $ is G-compact; so, for a suitable
sequence $g_{n}$, we have $g_{n}\mathbf{v}_{n}\rightarrow \mathbf{\bar{w}}%
\in \Gamma .$ We get the conclusion if we show that $V(\mathbf{u}%
_{n})\rightarrow 0.$ We have by (\ref{triplo}), that $d(\mathbf{u}_{n},%
\mathbf{v}_{n})\rightarrow 0$ and hence $d(g_{n}\mathbf{u}_{n},g_{n}\mathbf{v%
}_{n})\rightarrow 0$ and so, since $g_{n}\mathbf{v}_{n}\rightarrow \mathbf{%
\bar{w},}$ we have $g_{n}\mathbf{u}_{n}\rightarrow \mathbf{\bar{w}}\in
\Gamma .$ Therefore, by the continuity of $V$ and since $\mathbf{\bar{w}}\in
\Gamma ,$ we have $V\left( g_{n}\mathbf{u}_{n}\right) \rightarrow V\left( 
\mathbf{\bar{w}}\right) =0$ and we can conclude that $V\left( \mathbf{u}%
_{n}\right) \rightarrow 0.$

$\square$

\textbf{Proof of Th. \ref{astra1}. }By Theorem \ref{legicomp} for every $%
\delta \in \left( 0,\delta _{\infty }\right) $ $J_{\delta }$ is $G-$compact
and it has a minimizer $\mathbf{u}_{\delta }\neq 0$ with 
\begin{equation*}
E(\mathbf{u}_{\delta })=e_{\delta }
\end{equation*}%
where 
\begin{equation*}
e_{\delta }=\min \left\{ E(\mathbf{u}):\left\vert C(\mathbf{u})\right\vert
=c_{\delta }\right\} ,\text{ }c_{\delta }=\left\vert C(\mathbf{u}_{\delta
})\right\vert .
\end{equation*}%
So, in order to show that $\mathbf{u}_{\delta }$ is an hylomorphic soliton,
we need to show that 
\begin{equation*}
\Gamma \left( e_{\delta },c_{\delta }\right) =\left\{ \mathbf{u}\in X\ |\
\left\vert C(\mathbf{u})\right\vert =c_{\delta },E(\mathbf{u})=e_{\delta
}\right\}
\end{equation*}%
is $G$- compact and stable (see Definitions \ref{ds} and \ref{tdc}).

We set%
\begin{equation*}
V(\mathbf{u})=(E(\mathbf{u})-e_{\delta })^{2}+(\left\vert C(\mathbf{u}%
)\right\vert -c_{\delta })^{2}.
\end{equation*}%
Clearly 
\begin{equation*}
\Gamma \left( e_{\delta },c_{\delta }\right) =V^{-1}(0)
\end{equation*}
$V$ is $G$ compact. In fact:

Let $\mathbf{w}_{n}$ be a minimizing sequence for $V,$ then $V\left( \mathbf{%
w}_{n}\right) \rightarrow 0$ and consequently $E\left( \mathbf{w}_{n}\right)
\rightarrow e_{\delta }$ and $C\left( \mathbf{w}_{n}\right) \rightarrow
c_{\delta }$. Now, since 
\begin{equation*}
\min J_{\delta }=\frac{e_{\delta }}{c_{\delta }}+\delta e_{\delta },
\end{equation*}%
we have that $\mathbf{w}_{n}$ is a minimizing sequence also for $J_{\delta
}. $ Then, since by Theorem \ref{legicomp} $J_{\delta }$ is $G$-compact, we
get 
\begin{equation}
\mathbf{w}_{n}\ \text{is}\ G\text{-compact}.
\end{equation}%
So we conclude that $V$ is $G$-compact.

Then, by Lemma \ref{LC}, we deduce that $V^{-1}(0)$ $=\Gamma \left(
e_{\delta },c_{\delta }\right) $ is $G$ compact and that $V$ satisfies the
point (c) in Theorem \ref{propV}. Moreover $V$ satisfies also the points (a)
and (b) in Theorem \ref{propV}. So we conclude that $\Gamma \left( e_{\delta
},c_{\delta }\right) $ is stable.

$\square $

\textbf{Proof of Th. \ref{astra1+} }By Theorem \ref{astra1} for any $\delta
\in (0,\delta _{\infty })$ there exists a hylomorphic soliton $\mathbf{u}%
_{\delta }.$ By using Lemma \ref{BB}, we get different solitons for
different values of $\delta .$ Namely for $\delta _{1}<\delta _{2}$ we have $%
\Lambda (\mathbf{u}_{\delta _{1}})<\Lambda (\mathbf{u}_{\delta _{2}})$, $%
\left\vert C(\mathbf{u}_{\delta _{1}})\right\vert >\left\vert C(\mathbf{u}%
_{\delta _{2}})\right\vert $ and $E(\mathbf{u}_{\delta _{1}})>E(\mathbf{u}%
_{\delta _{2}}).$

$\square $

\section{The structure of hylomorphic solitons\label{mhs}}

\subsection{The meaning of the hylenic ratio}

Let $\left( X,\gamma \right) $ be a dynamical system of type FT. If $\mathbf{%
u}\in X$ is a finite energy field, usually it disperses as time goes on,
namely 
\begin{equation*}
\underset{t\rightarrow \infty }{\lim }\left\Vert \gamma _{t}\mathbf{u}%
\right\Vert _{\bigstar }=0.
\end{equation*}%
where 
\begin{equation*}
\left\Vert \mathbf{u}\right\Vert _{\bigstar }=\ \underset{x\in \mathbb{R}^{N}%
}{\sup }\int_{B_{1}(x)}\left\vert \mathbf{u}\right\vert _{V}\ dx,
\end{equation*}%
$V$ is the internal parameter space (cf. pag. \pageref{lilla}) and $%
B_{1}(x)= $ $\left\{ y\in \mathbb{R}^{N}:\left\vert x-y\right\vert
<1\right\} $. However, if the hylomorphy condition (\ref{hh}) is satisfied,
this dispersion in general does not occur. In fact we have the following
result:

\begin{proposition}
Assume that $X$ is compactly embedded into $L_{loc}^{1}\left( \mathbb{R}%
^{N},V\right) .$ Let $\mathbf{u}_{0}\in X$ such that $\Lambda \left( \mathbf{%
u}_{0}\right) <\Lambda _{0},\;$then 
\begin{equation*}
\underset{t\rightarrow \infty }{\min \lim }\left\Vert \mathbf{u}%
(t)\right\Vert _{\bigstar }>0
\end{equation*}%
where $\mathbf{u}(t)=\gamma _{t}\mathbf{u\ }$and $\gamma _{0}\mathbf{u=u}%
_{0} $.
\end{proposition}

\textbf{Proof}: Let $t_{n}\rightarrow \infty $ be a sequence of times such
that 
\begin{equation}
\underset{n\rightarrow \infty }{\lim }\ \left\Vert \mathbf{u}\left(
t_{n}\right) \right\Vert _{\bigstar }=\underset{t\rightarrow \infty }{\min
\lim }\left\Vert \mathbf{u}(t)\right\Vert _{\bigstar }.  \label{st}
\end{equation}

Since $\Lambda $ is a constant of motion 
\begin{equation*}
\Lambda \left( \mathbf{u}\left( t_{n}\right) \right) =\Lambda \left( \mathbf{%
u}_{0}\right) <\Lambda _{0}
\end{equation*}%
then, by the definition of $\Lambda _{0},$ may be taking a subsequence,
there is a sequence of translations $T_{x_{n}}$ such that 
\begin{equation}
T_{x_{n}}\mathbf{u}\left( t_{n}\right) =\mathbf{u}\left(
t_{n},x-x_{n}\right) =\mathbf{\bar{u}+w}_{n}  \label{p2}
\end{equation}%
where $\mathbf{\bar{u}}\neq 0$ and $\mathbf{w}_{n}\rightharpoonup 0$ in $X.$
Without loss of generality, we may assume that $\mathbf{\bar{u}}\neq 0$ in $%
B_{1}(0).$ Since $X$ is compactly embedded into $L_{loc}^{1}\left( \mathbb{R}%
^{N},V\right) ,$ we have that 
\begin{equation}
\int_{B_{1}(0)}\left\vert \mathbf{w}_{n}\right\vert _{V}\ dx\rightarrow 0.
\label{p3}
\end{equation}

By (\ref{p2}), we have that 
\begin{equation}
\left\vert T_{x_{n}}\mathbf{u}\left( t_{n}\right) \right\vert _{V}\geq
\left\vert \mathbf{\bar{u}}\right\vert _{V}\mathbf{-}\left\vert \mathbf{w}%
_{n}\right\vert _{V}.  \label{p4}
\end{equation}

Then, using (\ref{p4}), (\ref{p3}), we have that%
\begin{eqnarray*}
&&\underset{n\rightarrow \infty }{\min \lim }\int_{B_{1}(0)}\left\vert
T_{x_{n}}\mathbf{u}\left( t_{n}\right) \right\vert _{V}\ dx \\
&\geq &\ \underset{n\rightarrow \infty }{\lim }\left(
\int_{B_{1}(0)}\left\vert \mathbf{\bar{u}}\right\vert _{V}\ dx\ \mathbf{-}%
\int_{B_{1}(0)}\left\vert \mathbf{w}_{n}\right\vert _{V}\ dx\right) \\
&=&\int_{B_{1}(0)}\left\vert \mathbf{\bar{u}}\right\vert _{V}\ dx>0
\end{eqnarray*}%
Then%
\begin{equation}
\underset{n\rightarrow \infty }{\min \lim }\int_{B_{1}(0)}\left\vert
T_{x_{n}}\mathbf{u}\left( t_{n}\right) \right\vert _{V}\ dx>0.
\label{mentre}
\end{equation}

Finally, by (\ref{st}) and (\ref{mentre}), we get

\begin{eqnarray*}
\underset{}{\underset{t\rightarrow \infty }{\min \lim }}\ \left\Vert \mathbf{%
u}\left( t\right) \right\Vert _{\bigstar } &=&\underset{n\rightarrow \infty }%
{\lim }\ \left\Vert \mathbf{u}\left( t_{n}\right) \right\Vert _{\bigstar
}\geq \ \underset{n\rightarrow \infty }{\min \lim }\int_{_{B_{1}(x_{n})}}%
\left\vert \mathbf{u}\left( t_{n}\right) \right\vert _{V}\ dx \\
&=&\ \underset{n\rightarrow \infty }{\min \lim }\int_{B_{1}(0)}\left\vert
T_{x_{n}}\mathbf{u}\left( t_{n}\right) \right\vert _{V}\ dx>0.
\end{eqnarray*}

$\square$

Thus the hylomorphy condition prevents the dispersion. As we have seen in
the preceding section, (\ref{hh}) is also a fundamental assumption in
proving the existence of hylomorphic solitons.

\bigskip

Now, we assume $E$ and $C$ to be local quantities, namely, given $\mathbf{u}%
\in X,$ there exist the density functions $\rho _{E,\mathbf{u}}\left(
x\right) $ and $\rho _{C,\mathbf{u}}\left( x\right) \in L^{1}(\mathbb{R}%
^{N}) $ such that \ 
\begin{align*}
E\left( \mathbf{u}\right) & =\int \rho _{E,\mathbf{u}}\left( x\right) \ dx \\
C\left( \mathbf{u}\right) & =\int \rho _{C,\mathbf{u}}\left( x\right) \ dx
\end{align*}

Energy and hylenic densities $\rho _{E,\mathbf{u}},$ $\rho _{C,\mathbf{u}}$
allow to define the density of \textit{binding energy }as follows: 
\begin{equation}
\beta (t,x)=\beta _{\mathbf{u}}(t,x)=\left[ \rho _{E,\mathbf{u}}\left(
t,x\right) -\Lambda _{0}\cdot \left\vert \rho _{C,\mathbf{u}}\left(
t,x\right) \right\vert \right] ^{-}  \label{ben}
\end{equation}%
where $\left[ f\right] ^{-}$ denotes the negative part of $f.$

If $\mathbf{u}$ satisfies the hylomorphy condition, we have that $E\left( 
\mathbf{u}\right) <\Lambda _{0}\left\vert C\left( \mathbf{u}\right)
\right\vert $ and hence he have that $\beta _{\mathbf{u}}(t,x)\neq 0$ for
some $x\in \mathbb{R}^{N}.$

The support of the binding energy density is called \textit{bound matter
region; }more precisely we have the following definition

\begin{definition}
\label{supsol}Given any configuration $\mathbf{u}$, we define the \textbf{%
bound matter region} as follows 
\begin{equation*}
\Sigma \left( \mathbf{u}\right) =\overline{\left\{ x:\beta _{\mathbf{u}%
}(t,x)\neq 0\right\} }.
\end{equation*}%
If $\mathbf{u}_{0}$ is a soliton, the set $\Sigma \left( \mathbf{u}%
_{0}\right) $ is called \textbf{support of the soliton}\emph{\ }%
\index{soliton!support of the}at time\emph{\ }$t$.
\end{definition}

In the situation considered in this article, we will see that the solitons
satisfy the hylomorphy condition. Thus we may think that a soliton $\mathbf{u%
}_{0}$ consists of bound matter localized in a precise region of the space,
namely $\Sigma \left( \mathbf{u}_{0}\right) $. This fact gives the name to
this type of soliton from the Greek words \textquotedblright \textit{hyle}%
\textquotedblright =\textquotedblright \textit{matter}\textquotedblright\
and \textquotedblright \textit{morphe}\textquotedblright =\textquotedblright 
\textit{form}\textquotedblright .

\bigskip

\subsection{The \textit{swarm interpretation}\label{inter}}

Clearly the physical interpretation of hylomorphic solitons depends on the
model which we are considering. However we can always assume a \textit{%
conventional interpretations }which we will call \textit{swarm
interpretation }since the soliton is regarded as a \textit{swarm }of
particles bound together\textit{. }This iterpretation is consitent with the
model of the Q-ball.

We assume that $\mathbf{u}$ is a field which describes a fluid consisting of
particles; the particles density is given by the function $\rho
_{C}(t,x)=\rho _{C,\mathbf{u}}(t,x)$ which, of course satisfies a continuity
equation 
\begin{equation}
\partial _{t}\rho _{C}+\nabla \cdot \mathbf{J}_{C}=0  \label{CE}
\end{equation}%
where $\mathbf{J}_{C}$ is the flow of particles. Hence $C$ is the total
number of particles. Here the particles are not intended to be as in
"particle theory" but rather as in fluid dynamics, so that $C$ does not need
to be an integer number. Alternatively, if you like, you may think that $C$
is not the number of particles but it is proportional to it. Also, in some
equations as for example in NKG, $C$ can be negative; in this case, the
existence of \textit{antiparticles} is assumed.

Thus, the hylomorphy ratio 
\begin{equation*}
\Lambda \left( \mathbf{u}\right) =%
\frac{E\left( \mathbf{u}\right) }{\left\vert C\left( \mathbf{u}\right)
\right\vert }
\end{equation*}%
represents the average energy of each particle (or antiparticle). The number 
$\Lambda _{0}$ defined in (\ref{hylo}) is interpreted as the rest energy of
each particle when they do not interact with each other. If $\Lambda \left( 
\mathbf{u}\right) >\Lambda _{0},$ then the average energy of each particle
is bigger than the rest energy; if $\Lambda \left( \mathbf{u}\right)
<\Lambda _{0},$ the opposite occurs and this fact means that particles act
with each other with an attractive force.

If the particles were at rest and they were not acting on each other, their
energy density would be 
\begin{equation*}
\Lambda _{0}\cdot \left\vert \rho _{C}(t,x)\right\vert .
\end{equation*}%
If $\rho _{E}(t,x)$ denotes the energy density and if 
\begin{equation*}
\rho _{E}(t,x)<\Lambda _{0}\cdot \left\vert \rho _{C}(t,x)\right\vert ;
\end{equation*}%
then, in the point $x$ at time $t,$ the particles attract each other with a
force which is stronger than the repulsive forces; this explains the name 
\textit{density of} \textit{binding energy }given to $\beta (t,x)$ in (\ref%
{ben}).

Thus a soliton relative to the state $\mathbf{u}$ can be considered as a
"rigid" object occupying the region of space $\Sigma \left( \mathbf{u}%
\right) $ (cf. Def. \ref{supsol}); it consists of particles which stick to
each other; the energy to destroy the soliton is given by 
\begin{equation*}
\int \beta _{\mathbf{u}}(t,x)dx=\int_{\Sigma \left( \mathbf{u}\right)
}\left( \Lambda _{0}\left\vert \rho _{C}(t,x)\right\vert -\rho
_{E}(t,x)\right) dx.
\end{equation*}

\section{The Nonlinear Klein-Gordon-Maxwell equations}

Existence results of solitary waves for the Nonlinear Klein-Gordon-Maxwell
(NKGM) are stated in many papers (besides the papers quoted in the
introduction see also \cite{BF02}, \cite{ca}, \cite{dav}, \cite{tea}, \cite%
{tea2}, \cite{azzpomp}, \cite{azzpipomp}, \cite{befospin}, \cite{befo11max}, 
\cite{long06}). As stated in the introduction, in this section we prove the
existence of a \textit{continuous family} of \textit{stable} solitary waves
of NKGM.

\subsection{Basic features}

The nonlinear Klein-Gordon equation\ for a complex valued field $\psi ,$
defined on the space-time $\mathbb{R}^{4},$ can be written as follows:%
\begin{equation}
\square \psi +W^{\prime }(\psi )=0  \label{KG}
\end{equation}%
where 
\begin{equation*}
\square \psi =\frac{\partial ^{2}\psi }{\partial t^{2}}-\Delta \psi ,\;\;%
\text{\ }\Delta \psi =\frac{\partial ^{2}\psi }{\partial x_{1}^{2}}+\frac{%
\partial ^{2}\psi }{\partial x_{2}^{2}}+\frac{\partial ^{2}\psi }{\partial
x_{3}^{2}}
\end{equation*}%
and, with some abuse of notation, 
\begin{equation*}
W^{\prime }(\psi )=W^{\prime }(\left\vert \psi \right\vert )\frac{\psi }{%
\left\vert \psi \right\vert }
\end{equation*}%
for some smooth function $W:\left[ 0,\infty \right) \rightarrow \mathbb{R}.$
Hereafter $x=(x_{1},x_{2},x_{3})$ and $t$ will denote the space and time
variables. The field $\psi :$ $\mathbb{R}^{4}\rightarrow \mathbb{C}$ will be
called \textit{matter field}. If $W^{\prime }(s)$ is linear, $W^{\prime
}(s)=m^{2}s,$ $m\neq 0,$ equation (\ref{KG}) reduces to the Klein-Gordon
equation.

Consider the Abelian gauge theory in $\mathbb{R}^{4}$ equipped with the
Minkowski metric and described by the Lagrangian density (see e.g. \cite%
{befogranas}, \cite{yangL}, \cite{rub}) 
\begin{equation}
\mathcal{L}=\mathcal{L}_{0}+\mathcal{L}_{1}-W(\left\vert \psi \right\vert )
\label{marisa}
\end{equation}%
where 
\begin{equation*}
\mathcal{L}_{0}=\frac{1}{2}\left[ \left\vert D_{t}\psi \right\vert
^{2}-\left\vert \mathbf{D}_{x}\psi \right\vert ^{2}\right]
\end{equation*}%
\begin{equation*}
\mathcal{L}_{1}=\frac{1}{2}\left\vert \partial _{t}\mathbf{\QTR{mathbf}{A}}%
+\nabla \varphi \right\vert ^{2}-\frac{1}{2}\left\vert \nabla \times \mathbf{%
A}\right\vert ^{2}.
\end{equation*}%
Here $q$ denotes a positive parameter, $\nabla \times $ and $\nabla $ denote
respectively the curl and the gradient operators with respect to the $x$
variable, 
\begin{equation}
D_{t}=\frac{\partial }{\partial t}+iq\varphi ,\;D_{j}=\frac{\partial }{%
\partial x^{j}}-iqA_{j},\ \mathbf{D}_{x}\psi =(D_{1}\psi ,D_{2}\psi
,D_{3}\psi )  \label{cd}
\end{equation}%
are the covariant derivatives and finally $\varphi \in \mathbb{R}$ and $%
\mathbf{\QTR{mathbf}{A\mathbf{=(}}}A_{1},A_{2},A_{3}\mathbf{\QTR{mathbf}{)}}%
\in \mathbb{R}^{3}$ are the gauge potentials.

Now consider the total action 
\begin{equation}
\mathcal{S}=\int \left( \mathcal{L}_{0}+\mathcal{L}_{1}-W(\left\vert \psi
\right\vert )\right) dxdt.  \label{completa}
\end{equation}

Making the variation of $\mathcal{S}$ with respect to $\psi ,$ $\varphi $
and $\mathbf{A}$ we get the following system of equations 
\begin{equation}
D_{t}^{2}\psi -\mathbf{D}_{x}^{2}\psi +W^{\prime }(\psi )=0  \label{e1+}
\end{equation}%
\begin{equation}
\nabla \cdot \left( \frac{\partial \mathbf{\QTR{mathbf}{A}}}{\partial t}%
+\nabla \varphi \right) =q\left( \func{Im}\frac{\partial _{t}\psi }{\psi }%
+q\varphi \right) \left\vert \psi \right\vert ^{2}  \label{e2+}
\end{equation}%
\begin{equation}
\nabla \times \left( \nabla \times \mathbf{A}\right) +\frac{\partial }{%
\partial t}\left( \frac{\partial \mathbf{\QTR{mathbf}{A}}}{\partial t}%
+\nabla \varphi \right) =q\left( \func{Im}\frac{\nabla \psi }{\psi }-q%
\mathbf{A}\right) \left\vert \psi \right\vert ^{2}.  \label{e3+}
\end{equation}

The abelian gauge theory, namely equations (\ref{e1+}, \ref{e2+}, \ref{e3+}%
), provides a very elegant way to couple the Maxwell equation with matter if
we interpret $\psi $ as a matter field.

In order to give a more meaningful form to these equations, we will write $%
\psi $ in polar form 
\begin{equation*}
\psi (x,t)=u(x,t)\,e^{iS(x,t)},\;\;u\geq 0,\;\;S\in \mathbb{R}/2\pi \mathbb{Z%
}
\end{equation*}%
So (\ref{completa}) takes the following form 
\begin{align*}
\mathcal{S(}u,S,\varphi ,\mathbf{A})& =\int \int \left[ \frac{1}{2}\left( 
\frac{\partial u}{\partial t}\right) ^{2}-\frac{1}{2}\left\vert \nabla
u\right\vert ^{2}-W(u)\right] dxdt+ \\
& +\frac{1}{2}\int \int \left[ \left( \frac{\partial S}{\partial t}+q\varphi
\right) ^{2}-\left\vert \nabla S-q\mathbf{A}\right\vert ^{2}\right]
\,u^{2}dxdt \\
& +\frac{1}{2}\int \int \left( \left\vert \frac{\partial \mathbf{A}}{%
\partial t}\mathbf{+}\nabla \varphi \right\vert ^{2}-\left\vert \nabla
\times \mathbf{A}\right\vert ^{2}\right) dxdt
\end{align*}%
and the equations (\ref{e1+}, \ref{e2+}, \ref{e3+}) take the form: 
\begin{equation}
\square u+W^{\prime }(u)+\left[ \left\vert \nabla S-q\mathbf{A}\right\vert
^{2}-\left( \frac{\partial S}{\partial t}+q\varphi \right) ^{2}\right] \,u=0
\label{e1}
\end{equation}%
\begin{equation}
\frac{\partial }{\partial t}\left[ \left( \frac{\partial S}{\partial t}%
+q\varphi \right) u^{2}\right] -\nabla \cdot \left[ \left( \nabla S-q\mathbf{%
A}\right) u^{2}\right] =0  \label{e2}
\end{equation}%
\begin{equation}
\nabla \cdot \left( \frac{\partial \mathbf{\QTR{mathbf}{A}}}{\partial t}%
+\nabla \varphi \right) =q\left( \frac{\partial S}{\partial t}+q\varphi
\right) u^{2}\;  \label{e3}
\end{equation}%
\begin{equation}
\nabla \times \left( \nabla \times \mathbf{A}\right) +\frac{\partial }{%
\partial t}\left( \frac{\partial \mathbf{\QTR{mathbf}{A}}}{\partial t}%
+\nabla \varphi \right) =q\left( \nabla S-q\mathbf{A}\right) u^{2}.\;
\label{e4}
\end{equation}%
In order to show the relation of the above equations with the Maxwell
equations, we make the following change of variables: 
\begin{equation}
\mathbf{E=-}\left( \frac{\partial \mathbf{\QTR{mathbf}{A}}}{\partial t}%
+\nabla \varphi \right)  \label{pos1}
\end{equation}%
\begin{equation}
\mathbf{H}=\nabla \times \mathbf{A}  \label{pos2}
\end{equation}%
\begin{equation}
\rho =-\left( \frac{\partial S}{\partial t}+q\varphi \right) qu^{2}
\label{car}
\end{equation}%
\begin{equation}
\mathbf{j}=\left( \nabla S-q\mathbf{A}\right) qu^{2}.  \label{tar}
\end{equation}%
So (\ref{e3}) and (\ref{e4}) are the second couple of the Maxwell equations%
\index{equations!Maxwell} with respect to a matter distribution whose charge
and current density are respectively $\rho $ and $\mathbf{j}$: 
\begin{equation}
\nabla \cdot \mathbf{E}=\rho  \tag{\QTR{sc}{gauss}}  \label{gauss}
\end{equation}%
\begin{equation}
\nabla \times \mathbf{H}-%
\frac{\partial \mathbf{E}}{\partial t}=\mathbf{j}  \tag{\QTR{sc}{ampere}}
\label{ampere}
\end{equation}%
(\ref{pos1}) and (\ref{pos2}) give rise to the first couple of the Maxwell
equations 
\begin{equation}
\nabla \times \mathbf{E}+\frac{\partial \mathbf{H}}{\partial t}=0 
\tag{\QTR{sc}{faraday}}  \label{faraday}
\end{equation}%
\begin{equation}
\nabla \cdot \mathbf{H}=0.  \tag{\QTR{sc}{nomonopole}}  \label{monopole}
\end{equation}%
Using the variables $\mathbf{j}$ and $\rho ,$ equation (\ref{e1}) can be
written as follows 
\begin{equation}
\square u+W^{\prime }(u)+\frac{\mathbf{j}^{2}-\rho ^{2}}{q^{2}u}=0 
\tag{\QTR{sc}{matter}}  \label{materia}
\end{equation}%
and finally Equation (\ref{e2}) is the charge continuity equation%
\index{equation!continuity} 
\begin{equation}
\frac{\partial }{\partial t}\rho +\nabla \cdot \mathbf{j}=0.
\label{continuita}
\end{equation}

Notice that equation (\ref{continuita}) is a consequence of (\ref{gauss})
and (\ref{ampere}). In conclusion, an Abelian gauge theory, via equations (%
\ref{gauss},..,\ref{materia}), provides a model of interaction of the matter
field $\psi $ with the electromagnetic field $(\mathbf{E},\mathbf{H})$. In
fact that equations (\ref{gauss},..,\ref{materia}) are equivalent to (\ref%
{e1},..,\ref{e4}).

\subsection{Energy and charge\label{cl}}

Let examine the invariants of NKGM which are relevant for us, namely the
energy and the charge. In this subsection we compute these invariants using
the gauge invariant variables $u,\rho ,\mathbf{j},\mathbf{E},$ $\mathbf{H.}$

\textbf{Energy}. Energy, by definition, is the quantity which is preserved
by the time invariance of the Lagrangian. Using the gauge invariant
variables, the energy $E$ calculated along the solutions of equation (\ref%
{gauss}) takes the following form%
\begin{equation}
E=E_{m}+E_{f}  \label{sp11}
\end{equation}%
where 
\begin{equation*}
E_{m}=\frac{1}{2}\int \left[ \left( \frac{\partial u}{\partial t}\right)
^{2}+\left\vert \nabla u\right\vert ^{2}+W(u)+\frac{\rho ^{2}+\mathbf{j}^{2}%
}{2q^{2}u^{2}}\right] dx
\end{equation*}%
and%
\begin{equation*}
E_{f}=\frac{1}{2}\int \left( \mathbf{E}^{2}+\mathbf{H}^{2}\right) dx.
\end{equation*}

\textbf{Proof. }By the Noether's theorem (see e.g. \cite{Gelfand} or \cite%
{benci}), we have that, given the Lagrangian 
\begin{align*}
\mathcal{L}& =\frac{1}{2}\left( \frac{\partial u}{\partial t}\right) ^{2}-%
\frac{1}{2}\left\vert \nabla u\right\vert ^{2}-W(u)+ \\
& +\frac{1}{2}\left( \frac{\partial S}{\partial t}+q\varphi \right) ^{2}-%
\frac{1}{2}\left\vert \nabla S-q\mathbf{A}\right\vert ^{2}\,u^{2} \\
& +\frac{1}{2}\left\vert \frac{\partial \mathbf{A}}{\partial t}\mathbf{+}%
\nabla \varphi \right\vert ^{2}-\frac{1}{2}\left\vert \nabla \times \mathbf{A%
}\right\vert ^{2})
\end{align*}%
the density of energy takes the following form: 
\begin{equation*}
\frac{\partial \mathcal{L}}{\partial \left( \frac{\partial u}{\partial t}%
\right) }\cdot \frac{\partial u}{\partial t}+\frac{\partial \mathcal{L}}{%
\partial \left( \frac{\partial S}{\partial t}\right) }\cdot \frac{\partial S%
}{\partial t}+\frac{\partial \mathcal{L}}{\partial \left( \frac{\partial
\varphi }{\partial t}\right) }\cdot \frac{\partial \varphi }{\partial t}+%
\frac{\partial \mathcal{L}}{\partial \left( \frac{\partial \mathbf{A}}{%
\partial t}\right) }\cdot \frac{\partial \mathbf{A}}{\partial t}-\mathcal{L}
\end{equation*}%
Now we will compute each term. We have: 
\begin{equation}
\frac{\partial \mathcal{L}}{\partial \left( \frac{\partial u}{\partial t}%
\right) }\cdot \frac{\partial u}{\partial t}=\left( \frac{\partial u}{%
\partial t}\right) ^{2}  \label{l1}
\end{equation}%
\begin{eqnarray*}
\frac{\partial \mathcal{L}}{\partial \left( \frac{\partial S}{\partial t}%
\right) }\cdot \frac{\partial S}{\partial t} &=&\left( \frac{\partial S}{%
\partial t}+q\varphi \right) \frac{\partial S}{\partial t}\,u^{2} \\
&=&\left( \frac{\partial S}{\partial t}+q\varphi \right) \frac{\partial S}{%
\partial t}\,u^{2}+\left( \frac{\partial S}{\partial t}+q\varphi \right)
q\varphi u^{2}-\left( \frac{\partial S}{\partial t}+\varphi \right) q\varphi
u^{2} \\
&=&\left( \frac{\partial S}{\partial t}+q\varphi \right) ^{2}\,u^{2}-\left( 
\frac{\partial S}{\partial t}+q\varphi \right) q\varphi u^{2} \\
&=&\frac{\rho ^{2}}{q^{2}u^{2}}+\rho \varphi .
\end{eqnarray*}%
Multiplying by $\varphi $ equation (\ref{gauss}) and integrating, we get 
\begin{equation*}
-\int \mathbf{E\cdot }\nabla \varphi =\int \rho \varphi
\end{equation*}

Thus, replacing this expression in the above formula, we get 
\begin{equation}
\int \frac{\partial \mathcal{L}}{\partial \left( \frac{\partial S}{\partial t%
}\right) }\cdot \frac{\partial S}{\partial t}=\int \frac{\rho ^{2}}{%
q^{2}u^{2}}-\mathbf{E\cdot }\nabla \varphi  \label{l2}
\end{equation}%
Also we have 
\begin{equation}
\frac{\partial \mathcal{L}}{\partial \left( \frac{\partial \varphi }{%
\partial t}\right) }\cdot \frac{\partial \varphi }{\partial t}=0  \label{l3}
\end{equation}%
and 
\begin{equation}
\frac{\partial \mathcal{L}}{\partial \left( \frac{\partial \mathbf{A}}{%
\partial t}\right) }\cdot \frac{\partial \mathbf{A}}{\partial t}=\left( 
\frac{\partial \mathbf{A}}{\partial t}\mathbf{+}\nabla \varphi \right) \cdot 
\frac{\partial \mathbf{A}}{\partial t}=-\mathbf{E}\cdot \frac{\partial 
\mathbf{A}}{\partial t}  \label{l4}
\end{equation}%
Moreover, using the notation (\ref{pos1}, \ref{pos2}, \ref{car}, \ref{tar}),
we have that 
\begin{equation*}
\mathcal{L}=\frac{1}{2}\left( \frac{\partial u}{\partial t}\right) ^{2}-%
\frac{1}{2}\left\vert \nabla u\right\vert ^{2}-W(u)+\frac{\rho ^{2}-\mathbf{j%
}^{2}}{2q^{2}u^{2}}\,+\frac{\mathbf{E}^{2}-\mathbf{H}^{2}}{2}
\end{equation*}%
Then, by (\ref{l1},...,\ref{l4}) and the above expression for $\mathcal{L}$
we get 
\begin{eqnarray}
E\mathcal{(}u,S,\varphi ,\mathbf{A}) &=&\int \frac{\partial \mathcal{L}}{%
\partial \left( \frac{\partial u}{\partial t}\right) }\cdot \frac{\partial u%
}{\partial t}+\frac{\partial \mathcal{L}}{\partial \left( \frac{\partial S}{%
\partial t}\right) }\cdot \frac{\partial S}{\partial t}+\frac{\partial 
\mathcal{L}}{\partial \left( \frac{\partial \mathbf{A}}{\partial t}\right) }%
\cdot \frac{\partial \mathbf{A}}{\partial t}-\mathcal{L}  \notag \\
&=&\int \left( \frac{\partial u}{\partial t}\right) ^{2}+\frac{\rho ^{2}}{%
q^{2}u^{2}}-\mathbf{E\cdot }\nabla \varphi \mathbf{-E}\cdot \frac{\partial 
\mathbf{A}}{\partial t}-\mathcal{L}  \notag \\
&=&\int \left( \frac{\partial u}{\partial t}\right) ^{2}+\frac{\rho ^{2}}{%
q^{2}u^{2}}+\mathbf{E}^{2}  \label{en} \\
&&-\int \left[ \frac{1}{2}\left( \frac{\partial u}{\partial t}\right) ^{2}-%
\frac{1}{2}\left\vert \nabla u\right\vert ^{2}-W(u)+\frac{\rho ^{2}-\mathbf{j%
}^{2}}{2q^{2}u^{2}}\,+\frac{\mathbf{E}^{2}-\mathbf{H}^{2}}{2}\right]  \notag
\\
&=&\int \left[ \frac{1}{2}\left( \frac{\partial u}{\partial t}\right) ^{2}+%
\frac{1}{2}\left\vert \nabla u\right\vert ^{2}+W(u)+\frac{\rho ^{2}+\mathbf{j%
}^{2}}{2q^{2}u^{2}}\,+\frac{\mathbf{E}^{2}+\mathbf{H}^{2}}{2}\right]  \notag
\end{eqnarray}

$\square $

\bigskip

\textbf{Charge. }Using (\ref{continuita}), we see that the electric charge
has the following expression 
\begin{equation}
Q=\int \rho dx=-q\int (\partial _{t}S+q\varphi )u^{2}dx\;  \label{apr}
\end{equation}%
In order to be consistent with the previous literature (\cite{benci}, \cite%
{BBBM}, \cite{bbbs}, \cite{befoni}, \cite{befolak}, \cite{befo11max}, \cite%
{befospin}, \cite{befolib}), we will call \textit{charge }the following
quantity:\textit{\ } 
\begin{equation*}
C(\mathbf{u})=\frac{Q}{q}=-\int (\partial _{t}S+q\varphi )u^{2}dx.
\end{equation*}%
$C(\mathbf{u})$ is a dimensionless quantity which, in some interpretation of
NKGM, represents the number of particles (see \cite{Coleman86}, \cite{benci}%
, \cite{befolib}). In some of the quoted papers, $C(\mathbf{u})$ is called
hylenic charge and hence the ratio (\ref{lambda}) is called hylenic ratio.

\subsection{Existence of charged Q-balls%
\index{soliton!charged}}

We shall make the following assumptions on $W$:\label{pip}

\begin{itemize}
\item \label{Wi}\textbf{(W-0)} \textbf{(Positivity}) 
\begin{equation}
W(s)\geq 0;  \label{wi}
\end{equation}

\ 

\item \textbf{(W-i)} \textbf{(Nondegeneracy}) $W$ is a $C^{2}$ function s.t. 
$W(0)=W^{\prime }(0)=0\ $and 
\begin{equation}
W^{\prime \prime }(0)=m^{2}>0;  \label{wii}
\end{equation}

\ 

\item \textbf{(W-ii)} \textbf{(Hylomorphy}) if we set 
\begin{equation}
W(s)=%
\frac{1}{2}m^{2}s^{2}+N(s)  \label{NN}
\end{equation}%
then%
\begin{equation}
\exists s_{0}\in \mathbb{R}^{+}\text{ such that }N(s_{0})<0  \label{NNN}
\end{equation}

\item \textbf{(W-iii)} \textbf{(Growth condition}) 
\begin{equation}
|N^{\prime }(s)|\leq c_{1}s^{r-1}+c_{2}s^{q-1}\text{ for }q,\text{ }r\text{ }%
\in \text{ }(2,6)  \label{more}
\end{equation}
\end{itemize}

(W-0) implies that the energy $E$ in (\ref{sp11}) is positive; if this
condition does not hold, it is possible to have solitary waves, but not
hylomorphic solitons.

(W-i) In order to have solitary waves it is necessary to have $W^{\prime
\prime }(0)\geq 0.$ There are some results also when $W^{\prime \prime
}(0)=0 $ (null-mass case, see e.g. \cite{BL81} and \cite{BBR07}), however
the most interesting situations occur when $W^{\prime \prime }(0)>0.$

(W-ii) is the crucial assumption which characterizes the nonlinearity which
might produce hylomorphic solitons.\ 

The hylomorphy condition (W-ii) can also be written as follows: 
\begin{equation}
\alpha _{0}:=\ \underset{s\in \mathbb{R}^{+}}{\inf }\;\frac{W(s)}{\frac{1}{2}%
\left\vert s\right\vert ^{2}}<m^{2}  \label{wiii}
\end{equation}

(W-iii) if $W$ and hence $N$ is of class $C^{3},$ then (\ref{more}) reduces
to $|N^{\prime }(s)|\leq cs^{q-1}$ for $q<6,$ and this is the usual
subcritical growth condition.

Now we introduce the phase space $X.$

First observe that the term $\left( \rho ^{2}+\mathbf{j}^{2}\right) /u^{2}$
in (\ref{sp11}) is singular, so we introduce new gauge invariant variables
which eliminate this singularity:%
\begin{equation*}
\theta =\frac{\rho }{qu};\ \Theta =\frac{\mathbf{j}}{qu}.
\end{equation*}%
Using these new variables the energy takes the form:%
\begin{align*}
E\left( \mathbf{u}\right) & =\frac{1}{2}\int \left( \left\vert \partial
_{t}u\right\vert ^{2}+\left\vert \nabla u\right\vert ^{2}+\theta ^{2}+\Theta
^{2}+\mathbf{E}^{2}+\mathbf{H}^{2}\right) dx+\int W(u)dx \\
& =\frac{1}{2}\int \left[ \left\vert \partial _{t}u\right\vert
^{2}+\left\vert \nabla u\right\vert ^{2}+m^{2}u^{2}+\theta ^{2}+\Theta ^{2}+%
\mathbf{E}^{2}+\mathbf{H}^{2}\right] +\int N(u).
\end{align*}

The generic point in the phase space $X$ is given by%
\begin{equation*}
\mathbf{u}=\left( u,\hat{u},\theta ,\Theta ,\mathbf{E},\mathbf{H}\right)
\end{equation*}%
where $\hat{u}=\partial _{t}u$ is considered as independent variable; the
phase space is given by%
\begin{equation}
X=\left\{ \mathbf{u}\in \mathcal{H}:\nabla \cdot \mathbf{E}=q\theta u,\
\nabla \cdot \mathbf{H}=0\right\}  \label{xxx}
\end{equation}%
where $\mathcal{H}$ is the Hilbert space of the functions 
\begin{equation*}
\mathbf{u}=\left( u,\hat{u},\theta ,\Theta ,\mathbf{E},\mathbf{H}\right) \in
H^{1}\left( \mathbb{R}^{3}\right) \times L^{2}\left( \mathbb{R}^{3}\right)
^{11}
\end{equation*}%
equipped with the norm defined by the quadratic part of the energy, namely:%
\begin{equation}
\left\Vert \mathbf{u}\right\Vert ^{2}=\int \left[ \hat{u}^{2}+\left\vert
\nabla u\right\vert ^{2}+m^{2}u^{2}+\theta ^{2}+\Theta ^{2}+\mathbf{E}^{2}+%
\mathbf{H}^{2}\right] dx  \label{nor}
\end{equation}

In these new variables the energy and the hylenic charge become two
continuous functionals on $X$ having the form%
\begin{equation}
E\left( \mathbf{u}\right) =\frac{1}{2}\left\Vert \mathbf{u}\right\Vert
^{2}+\int N(u)dx  \label{placo}
\end{equation}

\begin{equation}
C\left( \mathbf{u}\right) =\int\theta u\ dx.  \label{plat}
\end{equation}

Our equations (\ref{materia}, \ref{gauss}, \ref{ampere}, \ref{faraday}, \ref%
{monopole}) become%
\begin{align}
\square u+W^{\prime }(u)+\frac{\Theta ^{2}-\theta ^{2}}{u}& =0  \notag \\
\nabla \cdot \mathbf{E}& =q\theta u  \notag \\
\nabla \times \mathbf{H}-\frac{\partial \mathbf{E}}{\partial t}& =q\Theta u
\label{equazioni} \\
\nabla \times \mathbf{E}+\frac{\partial \mathbf{H}}{\partial t}& =0  \notag
\\
\nabla \cdot \mathbf{H}& =0.  \notag
\end{align}

\begin{remark}
In the following we shall assume that the Cauchy problem for (NKGM) is well
posed in $X.$ Actually, in the literature there are few results relative to
this problem (we know only \cite{em}, \cite{kleinerman}, \cite{petre}) and
we do not know which are the assumptions that $W$ should satisfy. Also we
refer to \cite{befo11} for a discussion and some partial results on this
issue.
\end{remark}

We have the following existence results.

\begin{theorem}
\label{teoremnkgm} Assume that $W$ satisfies assumptions
(W-0),(W-i),(W-ii),(W-iii). Then there exists $\bar{q}>0$ such that for
every $q\in \left[ 0,\bar{q}\right] $ the dynamical system described by (\ref%
{equazioni}) has a family $\mathbf{u}_{\delta }$ ($\delta \in \left(
0,\delta _{\infty }\right) ,$ $\delta _{\infty }>0)$ of standing hylomorphic
solitons (Definition \ref{tdc}). Moreover if $\delta _{1}<\delta _{2}$ we
have that

\begin{itemize}
\item (a) $\Lambda (\mathbf{u}_{\delta _{1}})<\Lambda (\mathbf{u}_{\delta
_{2}})$

\item (b) $\left\vert C(\mathbf{u}_{\delta _{1}})\right\vert >\left\vert C(%
\mathbf{u}_{\delta 2})\right\vert $

\item (c) .$E(\mathbf{u}_{\delta _{1}})>E(\mathbf{u}_{\delta 2})$
\end{itemize}

\begin{theorem}
\label{imp}The solitons $\mathbf{u}_{\delta }=\left( u_{\delta },\hat{u}%
_{\delta },\theta _{\delta },\Theta _{\delta },\mathbf{E}_{\delta },\mathbf{H%
}_{\delta }\right) $ in Theorem \ref{teoremnkgm} are stationary solutions of
(\ref{equazioni}), this means that $\hat{u}_{\delta }=\Theta _{\delta }=%
\mathbf{H}_{\delta }=0,$ $\mathbf{E}_{\delta }=-\nabla \varphi _{\delta }$
and $u_{\delta },\theta _{\delta },\varphi _{\delta }$ solve the equations 
\begin{align}
-\Delta u_{\delta }+W^{\prime }(u_{\delta })-\frac{\theta _{\delta }^{2}}{%
u_{\delta }}& =0  \label{stat1} \\
-\Delta \varphi _{\delta }& =-q\theta _{\delta }u_{\delta }.  \label{stat2}
\end{align}
\end{theorem}
\end{theorem}

We shall prove Theorem \ref{teoremnkgm} by using the abstract Theorem \ref%
{astra1+}. First of all observe that the energy and the hylenic charge $E$
and $C$, defined in (\ref{placo}) and (\ref{plat}) are invariant under
translations i. e. under the action of the group $\mathcal{T}$ defined in (%
\ref{gg}).

We shall see that assumptions (\ref{wi}),...,(\ref{more}) on $W$ permit to
show that assumptions (EC-0), (EC-1), (EC-2), (EC-3), (\ref{poo}) and (\ref%
{hh}) of the abstract theorem \ref{astra1+} are satisfied.

The next two lemmas, whose proofs follow standard arguments, state that $E$
satisfies the coercivity assumption (EC-3) and that both $E$ and $C$ satisfy
the splitting property (EC-2).

\begin{lemma}
\label{coerciv} Let the assumptions of Theorem \ref{teoremnkgm} be
satisfied, then $E$ defined by (\ref{placo}) satisfies (EC-3), namely for
any sequence $\mathbf{u}_{n}$ $=\left( u_{n},\hat{u}_{n},\theta _{n},\Theta
_{n},\mathbf{E}_{n},\mathbf{H}_{n}\right) $ in $\mathcal{H}$ such that $E(%
\mathbf{u}_{n})\rightarrow 0$ (respectively $E(\mathbf{u}_{n})$ bounded)$,$
we have $\left\Vert \mathbf{u}_{n}\right\Vert \rightarrow 0$ (respectively $%
\left\Vert \mathbf{u}_{n}\right\Vert $ bounded)$,$ where $\left\Vert \mathbf{%
\cdot }\right\Vert $ is defined in (\ref{nor}).
\end{lemma}

\textbf{Proof. }See proof of Lemma 23 in \cite{befo11max}.

$\square $

\begin{lemma}
\label{split copy(1)} Let the assumptions of Theorem \ref{teoremnkgm} be
satisfied, then $E$ and $C$ satisfy the splitting property (EC-2).
\end{lemma}

\textbf{Proof.} See proof of Lemma 22 in \cite{befo11max}.

$\square $

It remains to prove that the hylomorphy condition (\ref{hh}) holds.

\subsection{Analysis of the hylenic ratio%
\index{hylenic ratio!for NKGM}}

First of all we set:%
\begin{equation}
\left\Vert \mathbf{u}\right\Vert _{\sharp }=\left\Vert \left( u,%
\hat{u},\theta ,\Theta ,\mathbf{E},\mathbf{H}\right) \right\Vert _{\sharp
}=\max \left( \left\Vert u\right\Vert _{L^{r}},\left\Vert u\right\Vert
_{L^{q}}\right)  \label{semi}
\end{equation}

where $r,$ $q$ are introduced in (\ref{more}). With some abuse of notation
we shall write $\max \left( \left\Vert u\right\Vert _{L^{r}},\left\Vert
u\right\Vert _{L^{q}}\right) =\left\Vert u\right\Vert _{\sharp }$.\ 

\begin{lemma}
\label{nonvanishing4}\textit{\textbf{\ }}The seminorm $\left\Vert \mathbf{u}%
\right\Vert _{\sharp }$ defined in (\ref{semi}) satisfies the property (\ref%
{seminorm}), namely%
\begin{equation*}
\left\{ \mathbf{u}_{n}\ \text{is a }\mathcal{T-}\text{vanishing sequence}%
\right\} \Rightarrow \left\Vert u_{n}\right\Vert _{\sharp }\rightarrow 0.
\end{equation*}%
where $\mathcal{T}$ is defined in (\ref{gg}).
\end{lemma}

\textbf{Proof. } Let $u_{n}\ $be a bounded sequence in $H^{1}\left( \mathbb{R%
}^{3}\right) $ 
\begin{equation}
\left\Vert u_{n}\right\Vert _{H^{1}(\mathbb{R}^{3})}^{2}\leq M  \label{impl}
\end{equation}

such that, up to a subsequence, 
\begin{equation}
\left\Vert u_{n}\right\Vert _{\sharp }\geq a>0.  \label{implo}
\end{equation}

$.$ We need to show that $u_{n}$ is not $\mathcal{T-}$ vanishing.

May be taking a subsequence, we have that at least one of the following
holds:

\begin{itemize}
\item (i) $\left\Vert u_{n}\right\Vert _{\sharp }=\left\Vert
u_{n}\right\Vert _{L^{r}}$

\item (ii) $\left\Vert u_{n}\right\Vert _{\sharp }=\left\Vert
u_{n}\right\Vert _{L^{q}}$
\end{itemize}

Now suppose that (i) holds (If (ii) holds, we will argue in the same way
replacing $r$ with $q).$

We set for $j\in \mathbb{Z}^{3}$ 
\begin{equation*}
Q_{j}=j+Q=\left\{ j+q:q\in Q\right\}
\end{equation*}%
where $Q$ is now the cube defined as follows 
\begin{equation*}
Q=\left\{ \left( x_{1},..,x_{n}\right) \in \mathbb{R}^{3}:0\leq
x_{i}<1\right\} \text{.}
\end{equation*}%
Clearly 
\begin{equation*}
\mathbb{R}^{3}=\dbigcup\limits_{j}Q_{j}.
\end{equation*}%
Now let $c$ be the constant for the Sobolev embedding $H^{1}\left(
Q_{j}\right) \subset L^{t}\left( Q_{j}\right) .$ We have 
\begin{align*}
0& <a^{r}\leq \int \left\vert u_{n}\right\vert
^{r}=\sum_{j}\int_{Q_{j}}\left\vert u_{n}\right\vert ^{r}=\sum_{j}\left\Vert
u_{n}\right\Vert _{L^{t}\left( Q_{j}\right) }^{r-2}\left\Vert
u_{n}\right\Vert _{L^{t}\left( Q_{j}\right) }^{2} \\
& \leq \ \left( \underset{j}{\sup }\left\Vert u_{n}\right\Vert _{L^{t}\left(
Q_{j}\right) }^{r-2}\right) \cdot \sum_{j}\left\Vert u_{n}\right\Vert
_{L^{r}\left( Q_{j}\right) }^{2} \\
& \leq \ c\left( \underset{j}{\sup }\left\Vert u_{n}\right\Vert
_{L^{t}\left( Q_{j}\right) }^{r-2}\right) \cdot \sum_{j}\left\Vert
u_{n}\right\Vert _{H^{1}\left( Q_{j}\right) }^{2} \\
& =c\left( \underset{j}{\sup }\left\Vert u_{n}\right\Vert _{L^{r}\left(
Q_{j}\right) }^{r-2}\right) \left\Vert u_{n}\right\Vert _{H^{1}}^{2}\leq
cM\left( \underset{j}{\sup }\left\Vert u_{n}\right\Vert _{L^{t}\left(
Q_{j}\right) }^{r-2}\right) .
\end{align*}%
where $M$ and $a$ are the constants respectively in (\ref{impl}) and (\ref%
{implo}). Then%
\begin{equation*}
\left( \underset{j}{\sup }\left\Vert u_{n}\right\Vert _{L^{t}\left(
Q_{j}\right) }\right) \geq \left( \frac{a^{t}}{cM}\right) ^{1/(t-2)}
\end{equation*}

Then, for any $n,$ there exists $j_{n}\in \mathbb{Z}^{3}$ such that 
\begin{equation}
\left\Vert u_{n}\right\Vert _{L^{r}\left( Q_{j_{n}}\right) }\geq \alpha >0.
\label{caca}
\end{equation}%
Then, since $\left( T_{j_{n}}u_{n}\right) (x)=u_{n}(x+j_{n})$ (see ( \ref%
{ggg})), we have 
\begin{equation}
\left\Vert T_{j_{n}}u_{n}\right\Vert _{L^{r}(Q_{0})}=\left\Vert
u_{n}\right\Vert _{L^{r}(Q_{j_{n}})}\geq \alpha >0.  \label{chicco}
\end{equation}

Since $u_{n}$ is bounded, also $T_{j_{n}}u_{n}$ is bounded in $H^{1}(\mathbb{%
R}^{3}).$ Then we have, up to a subsequence, that $T_{j_{n}}u_{n}%
\rightharpoonup u_{0}$ weakly in $H^{1}(\mathbb{R}^{3})$ and hence strongly
in $L^{r}(Q)$. By (\ref{chicco}), $u_{0}\neq 0.$

So we conclude that $u_{n}$ is not $\mathcal{T-}$ vanishing.

$\square $

Now, as usual, we set%
\begin{equation*}
\Lambda _{0}:=\ \inf \left\{ \lim \inf \ \Lambda (\mathbf{u}_{n})\ |\ 
\mathbf{u}_{n}\ \text{is a }\mathcal{T-}\text{vanishing sequence}\right\}
\end{equation*}

\begin{eqnarray}
\Lambda _{\sharp } &=&\ \underset{\left\Vert \mathbf{u}\right\Vert _{\sharp
}\rightarrow 0}{\lim \inf }\Lambda (\mathbf{u})=\   \label{lino} \\
&&\underset{\varepsilon \rightarrow 0}{\lim }\ \inf \left\{ \Lambda
(\varepsilon u,\hat{u},\theta ,\Theta ,\mathbf{E},\mathbf{H})\ |u\in H^{1},\
(\hat{u},\theta ,\Theta ,\mathbf{E},\mathbf{H)}\in \left( L^{2}\right)
^{11};\ \left\Vert u\right\Vert _{\sharp }=1\right\} .  \notag
\end{eqnarray}

By the definition of $\Lambda _{0}$ and $\Lambda _{\sharp }$ and lemma \ref%
{nonvanishing4}, we have that%
\begin{equation}
\Lambda _{0}\geq \Lambda _{\sharp }  \label{la}
\end{equation}

The following lemma holds:

\begin{lemma}
\label{preparatorio3}Let $W$ satisfy assumption (\ref{more}), then the
following inequality holds 
\begin{equation}
\Lambda _{\sharp }\geq m.  \label{le}
\end{equation}
\end{lemma}

\textbf{Proof.} First of all observe that by (\ref{more}) we have%
\begin{align*}
\left\vert \int N(\left\vert u\right\vert )dx\right\vert & \leq k_{1}\int
\left\vert u\right\vert ^{r}+k_{2}\int \left\vert u\right\vert ^{q}\  \\
& \leq k_{1}\left\Vert u\right\Vert _{\sharp }^{r}+k_{2}\left\Vert
u\right\Vert _{\sharp }^{q}.
\end{align*}%
So, if we take $\left\Vert u\right\Vert _{\sharp }=1$ and $\varepsilon >0,$%
we get%
\begin{equation}
\left\vert \int N(\left\vert \varepsilon u\right\vert )dx\right\vert \leq
k_{1}\varepsilon ^{r}+k_{2}\varepsilon ^{q}.  \label{co}
\end{equation}

By the Sobolev embeddings, there is $k_{3}>0$ such that 
\begin{equation}
\int \left( \left\vert \nabla u\right\vert ^{2}+m^{2}u^{2}\right) dx\geq
k_{3}\left\Vert u\right\Vert _{\sharp }^{2}\text{ }  \label{cot}
\end{equation}%
Now, choose 
\begin{equation*}
2<s<\min (r,q).
\end{equation*}%
Since $r,q>s$, we have, by (\ref{cot}), (\ref{co}) and taking $\varepsilon
>0 $ small enough, that 
\begin{align*}
& \varepsilon ^{s}\int \left( \left\vert \nabla u\right\vert
^{2}+m^{2}u^{2}\right) dx-\left\vert \int N(\left\vert \varepsilon
u\right\vert )dx\right\vert \\
& \geq \varepsilon ^{s}k_{3}\left\Vert \psi \right\Vert _{\sharp
}^{2}-k_{1}\varepsilon ^{r}-k_{2}\varepsilon ^{q}=k_{3}\varepsilon
^{s}-k_{1}\varepsilon ^{r}-k_{2}\varepsilon ^{q}\geq 0
\end{align*}%
So 
\begin{equation}
\left\vert \int N(\varepsilon \left\vert u\right\vert )dx\right\vert \leq
\varepsilon ^{s}\int \left( \left\vert \nabla u\right\vert
^{2}+m^{2}u^{2}\right) dx\text{ where }s>2.  \label{cinzia3}
\end{equation}%
Then, by using (\ref{cinzia3}), for any $\mathbf{u}=\left( u,\hat{u},\theta
,\Theta ,\mathbf{E},\mathbf{H}\right) ,$ with $u\in H^{1},\ \left\Vert
u\right\Vert _{\sharp }=1$ and any $(\hat{u},\theta ,\Theta ,\mathbf{E},%
\mathbf{H)}\in \left( L^{2}\right) ^{11},$ we have, for $\varepsilon >0$
small

\begin{align*}
\Lambda (\varepsilon u,\hat{u},\theta ,\Theta ,\mathbf{E},\mathbf{H})& \geq 
\frac{\frac{1}{2}\int \left( \theta ^{2}+\varepsilon ^{2}\left\vert \nabla
u\right\vert ^{2}+\varepsilon ^{2}m^{2}\left\vert u\right\vert ^{2}\right)
dx+\int N(\left\vert \varepsilon u\right\vert )dx}{\varepsilon \left\vert
\int \theta u\right\vert } \\
& \geq \frac{\frac{1}{2}\int \theta ^{2}+\left( \frac{\varepsilon ^{2}}{2}%
-\varepsilon ^{s}\right) \int \left( \left\vert \nabla u\right\vert
^{2}+m^{2}\left\vert u\right\vert ^{2}\right) }{\varepsilon \left\Vert
\theta \right\Vert _{L^{2}}\left\Vert u\right\Vert _{L^{2}}} \\
& \geq \frac{\left( \int \theta ^{2}dx\right) ^{1/2}\cdot \varepsilon m\sqrt{%
1-2\varepsilon ^{s-2}}\left( \int u^{2}dx\right) ^{1/2}}{\varepsilon
\left\Vert \theta \right\Vert _{L^{2}}\left\Vert u\right\Vert _{L^{2}}}=m%
\sqrt{1-2\varepsilon ^{s-2}}.
\end{align*}%
Then, since $s>2,$ we have 
\begin{equation}
\underset{\varepsilon \rightarrow 0}{\Lambda _{\sharp }=\lim }\ \inf \left\{
\Lambda (\varepsilon u,\hat{u},\theta ,\Theta ,\mathbf{E},\mathbf{H})\ |u\in
H^{1},\ (\hat{u},\theta ,\Theta ,\mathbf{E},\mathbf{H)}\in \left(
L^{2}\right) ^{11};\ \left\Vert u\right\Vert _{\sharp }=1\right\} \geq m
\label{nuovo}
\end{equation}

$\square$

Next we will show that the hylomorphy assumption (\ref{hh}) is satisfied.

\begin{lemma}
\label{ve}Assume that $W$ satisfies (W-0),...,(W-iii) and (\ref{more}) then%
\begin{equation}
\underset{\mathbf{u}\in X}{\inf }\Lambda (\mathbf{u})<\Lambda _{0}.
\label{li}
\end{equation}
\end{lemma}

\textbf{Proof}. We shall prove that%
\begin{equation}
\underset{}{\Lambda _{\ast }=\underset{\mathbf{u}\in X}{\inf }\Lambda (%
\mathbf{u})}<m  \label{ing}
\end{equation}%
So (\ref{li}) will follow from (\ref{la}) and (\ref{le}) and (\ref{ing}).

Let $R>0;$ set 
\begin{equation}
u_{R}=\left\{ 
\begin{array}{cc}
s_{0} & if\;\;|x|<R \\ 
0 & if\;\;|x|>R+1 \\ 
\frac{|x|}{R}s_{0}-(\left\vert x\right\vert -R)\frac{R+1}{R}s_{0} & 
if\;\;R<|x|<R+1%
\end{array}%
\right. .  \label{inff}
\end{equation}%
where $R>1.$

By the hylomorphy assumption (\ref{wii}) there exist $\alpha \in (0,m)$ such
that 
\begin{equation}
W(s_{0})\leq \frac{1}{2}\alpha ^{2}s_{0}^{2}  \label{pelo}
\end{equation}

Now let $\varphi _{R}\in \mathcal{D}^{1,2}$ denote the solution of the
following equation%
\begin{equation}
\Delta \varphi =-q\alpha u_{R}^{2}.  \label{mar}
\end{equation}

We have%
\begin{align*}
\Lambda_{\ast} & =\ \underset{\mathbf{u}\in X}{\inf}\ \frac{\frac{1}{2}%
\left\Vert \mathbf{u}\right\Vert ^{2}+\int N(u)dx}{\left\vert C\left( 
\mathbf{u}\right) \right\vert } \\
& =\ \underset{\mathbf{u}\in X}{\inf}\frac{\frac{1}{2}\int\left[ \hat{u}%
^{2}+\left\vert \nabla u\right\vert ^{2}+\theta^{2}+\Theta^{2}+\mathbf{E}%
^{2}+\mathbf{H}^{2}\right] dx+\int W(u)dx}{\left\vert \int\theta u\
dx\right\vert }.
\end{align*}

Now remember that $\mathbf{u}=\left( u,\hat{u},\theta ,\Theta ,\mathbf{E},%
\mathbf{H}\right) $ and take $\mathbf{u}=\mathbf{u}_{R}$ with 
\begin{equation*}
\mathbf{u}_{R}=\left( u_{R},0,\alpha u_{R},0,\nabla \varphi _{R},\mathbf{0}%
\right) .
\end{equation*}%
By (\ref{mar}), $\mathbf{u}_{R}\in X;$ then we have

\begin{align}
\Lambda _{\ast }& =\ \underset{\mathbf{u}\in X}{\inf }\ \frac{\frac{1}{2}%
\left\Vert \mathbf{u}\right\Vert ^{2}+\int N(u)dx}{\left\vert C\left( 
\mathbf{u}\right) \right\vert }\leq \frac{\frac{1}{2}\left\Vert \mathbf{u}%
_{R}\right\Vert ^{2}+\int N(u_{R})dx}{\left\vert C\left( \mathbf{u}%
_{R}\right) \right\vert }  \notag \\
& =\frac{\frac{1}{2}\int \left[ \left\vert \nabla u_{R}\right\vert
^{2}+\alpha ^{2}u_{R}^{2}+\left\vert \nabla \varphi _{R}\right\vert ^{2}%
\right] dx+\int W(u_{R})dx}{\alpha \int u_{R}^{2}\ dx}  \notag \\
& \leq \frac{\frac{1}{2}\int_{\left\vert x\right\vert <R}\left[ \left\vert
\nabla u_{R}\right\vert ^{2}+\alpha ^{2}u_{R}^{2}\right] +\int_{\left\vert
x\right\vert <R}W(u_{R})}{\alpha \int_{\left\vert x\right\vert <R}u_{R}^{2}\
dx}  \notag \\
& +\frac{\frac{1}{2}\int_{R<\left\vert x\right\vert <R+1}\left[ \left\vert
\nabla u_{R}\right\vert ^{2}+\alpha ^{2}u_{R}^{2}\right] +\int_{R<\left\vert
x\right\vert <R+1}W(u_{R})}{\alpha \int_{\left\vert x\right\vert
<R}u_{R}^{2}\ }+\frac{\frac{1}{2}\int \left\vert \nabla \varphi
_{R}\right\vert ^{2}}{\alpha \int_{\left\vert x\right\vert <R}u_{R}^{2}\ } 
\notag \\
& =\frac{\frac{1}{2}\int_{\left\vert x\right\vert <R}\alpha
^{2}s_{0}^{2}+\int_{\left\vert x\right\vert <R}W(s_{0})}{\alpha
\int_{\left\vert x\right\vert <R}s_{0}^{2}}+\frac{c_{1}R^{2}}{\alpha
\int_{\left\vert x\right\vert <R}s_{0}^{2}}+\frac{\frac{1}{2}\int \left\vert
\nabla \varphi _{R}\right\vert ^{2}}{\alpha \int_{\left\vert x\right\vert
<R}s_{0}^{2}\ \ }  \notag \\
& \leq \alpha +\frac{c_{2}}{\alpha R}+\frac{\frac{1}{2}\int \left\vert
\nabla \varphi _{R}\right\vert ^{2}}{\frac{4}{3}\pi \alpha s_{0}^{2}R^{3}\ }
\label{star}
\end{align}%
where the last inequality is a consequence of (\ref{pelo}).

In order to estimate the term containing $\varphi _{R}$ in (\ref{star}), we
remember that $\varphi _{R}$ is the solution of (\ref{mar}). Observe that $%
u_{R}^{2}$ has radial symmetry and that the electric field outside any
spherically symmetric charge distribution is the same as if all of the
charge were concentrated into a point. So $\left\vert \nabla \varphi
_{R}\left( r\right) \right\vert $ corresponds to the strength of an
electrostatic field at distance $r,$ created by an electric charge given by 
\begin{equation*}
\left\vert C_{el}\right\vert =\dint\limits_{\left\vert x\right\vert \leq
r}q\alpha u_{R}^{2}dx=4\pi \dint\limits_{0}^{r}q\alpha u_{R}^{2}v^{2}dv
\end{equation*}%
and located at the origin. So we have%
\begin{equation*}
\left\vert \nabla \varphi _{R}\left( r\right) \right\vert =\frac{\left\vert
C_{el}\right\vert }{r^{2}}\left\{ 
\begin{array}{cc}
=\frac{4}{3}\pi q\alpha s_{0}^{2}r & if\ r<R \\ 
\leq \frac{4}{3}\pi q\alpha s_{0}^{2}\frac{(R+1)^{3}}{r^{2}} & if\ r\geq R%
\end{array}%
\right.
\end{equation*}

Then%
\begin{align*}
\int \left\vert \nabla \varphi _{R}\right\vert ^{2}dx& \leq c_{3}q^{2}\alpha
^{2}s_{0}^{4}\left( \int_{r<R}r^{4}dr+\int_{r>R}\frac{(R+1)^{6}}{r^{2}}%
dr\right) \\
& \leq c_{4}q^{2}\alpha ^{2}s_{0}^{4}\left( R^{5}+\frac{(R+1)^{6}}{R}\right)
\leq c_{5}q^{2}\alpha ^{2}s_{0}^{4}R^{5}.
\end{align*}%
Then%
\begin{equation}
\frac{\frac{1}{2}\int \left\vert \nabla \varphi _{R}\right\vert ^{2}}{\frac{4%
}{3}\pi \alpha s_{0}^{2}R^{3}\ }\leq c_{6}q^{2}\alpha s_{0}^{2}R^{2}.
\end{equation}%
By (\ref{star}), we get

\begin{equation}
\Lambda _{\ast }\leq \alpha +\frac{c_{1}}{\alpha R}+c_{6}q^{2}\alpha
s_{0}^{2}R^{2}.  \label{quasi}
\end{equation}

Now set 
\begin{equation*}
m-\alpha =2\varepsilon
\end{equation*}%
and take 
\begin{equation*}
R=\frac{c_{1}}{\alpha \varepsilon },\text{ }0<q<\sqrt{\frac{\varepsilon
^{3}\alpha }{s_{0}^{2}c_{1}^{2}c_{6}}}.
\end{equation*}%
With these choices of $R$ and $q,$ a direct calculation shows that%
\begin{equation}
\alpha +\frac{c_{1}}{\alpha R}+c_{6}q^{2}\alpha s_{0}^{2}R^{2}<m.
\label{cucu}
\end{equation}%
Then, by (\ref{quasi}) and (\ref{cucu}), we get that there exists a positive
constant $c$ such that, for $0<q<\frac{c}{\bar{s}}\sqrt{\left( m-\alpha
\right) ^{3}\alpha }$, we have 
\begin{equation}
\underset{\mathbf{u}\in X}{\inf }\Lambda (\mathbf{u})<m.  \label{dopo}
\end{equation}

$\square $

.

\textbf{Proof of Th}.\ref{teoremnkgm} Assumptions (EC-0), (EC-1) of Theorem%
\ref{astra1+} are clearly satisfied. By Lemma \ref{split copy(1)} $E$ and $C$
satisfy the splitting property (EC-2) . . By Lemma \ref{coerciv} and by
Lemma \ref{ve} also the coercitivity assumption (EC-3) and hylomorphy
condition (\ref{hh}) are satisfied.

Finally it remains to show that also (\ref{poo}) is satisfied. To this end
let 
\begin{equation*}
\mathbf{u}=\left( u,\hat{u},\theta ,\Theta ,\mathbf{E},\mathbf{H}\right) \in
H^{1}\left( \mathbb{R}^{3}\right) \times L^{2}\left( \mathbb{R}^{3}\right)
^{11}
\end{equation*}%
be a solution of $E^{\prime }(\mathbf{u})=0,$ then it is easy to see that $%
\left( \hat{u},\theta ,\Theta ,\mathbf{E},\mathbf{H}\right) \mathbf{=}0$ and 
$u\in H^{1}\left( \mathbb{R}^{3}\right) $ solves the equation 
\begin{equation*}
-\Delta u+W^{\prime }(u)=0.
\end{equation*}%
So, since $W\geq 0,$ we have by the Derrick-Pohozaev identity \cite{der}
that also $u=0.$We conclude that 
\begin{equation*}
\mathbf{u}=\left( u,\hat{u},\theta ,\Theta ,\mathbf{E},\mathbf{H}\right) =0.
\end{equation*}

So all the assumptions of the Theorem \ref{astra1+} are satisfied and the
conclusion follows.

$\square $.

\textbf{Proof of Th. \ref{imp}. } Let%
\begin{equation*}
\mathbf{u}_{\delta }=\left( u_{\delta },0,\theta _{\delta },0,\mathbf{E}%
_{\delta },\mathbf{0}\right) \in X=\left\{ \mathbf{u}\in \mathcal{H}:\nabla
\cdot \mathbf{E}=q\theta u,\ \nabla \cdot \mathbf{H}=0\right\}
\end{equation*}%
be as in Theorem \ref{imp}.

So $\mathbf{u}_{\delta }$ minimizes the energy $E$ (see (\ref{placo})) on
the manifold%
\begin{equation*}
X_{\delta }=\left\{ \mathbf{u}\in X:C(\mathbf{u)}=C(\mathbf{u}_{\delta
})=\sigma _{\delta }\right\} .
\end{equation*}%
If we write $\mathbf{E}=-\nabla \varphi $, the constraint $\nabla \cdot 
\mathbf{E}=q\theta u$ becomes 
\begin{equation*}
-\Delta \varphi =q\theta u.
\end{equation*}%
So $\mathbf{u}_{\delta }$ is a critical point of $E$ on the manifold (in $%
\mathcal{H)}$ made up by those $\mathbf{u}=$ $\left( u,0,\theta ,0,\nabla
\varphi ,\mathbf{0}\right) $ satisfying the constraints%
\begin{equation}
\Delta \varphi =q\theta u  \label{vic1}
\end{equation}

\begin{equation}
C(\mathbf{u)=}\int\theta u\ dx=\sigma_{\delta}.  \label{vic2}
\end{equation}

Therefore, for suitable Lagrange multipliers $\lambda \in \mathbb{R},$ $\xi
\in \mathcal{D}^{1,2}$ ($\mathcal{D}^{1,2}$ is the closure of $C_{0}^{\infty
}$ with respect to the norm $\left\Vert \nabla \varphi \right\Vert _{L^{2}}$%
), we have that $\mathbf{u}_{\delta }$ is a critical point of 
\begin{equation}
E_{\lambda ,\xi }(\mathbf{u)}=E(\mathbf{u)+}\lambda \left( \int \theta u\ dx%
\mathbf{-}\sigma _{\delta }\right) +\left\langle \xi ,-\Delta \varphi
+q\theta u\right\rangle  \label{lag}
\end{equation}%
where $\left\langle \ \cdot \ ,\ \cdot \ \right\rangle $ denotes the duality
map in $\mathcal{D}^{1,2}.$ It is easy to show that $E_{\lambda ,\xi
}^{\prime }(\mathbf{u}_{\delta })=0$ gives the equations%
\begin{align}
-\Delta u_{\delta }+W^{\prime }(u_{\delta })+\lambda \theta _{\delta }+q\xi
\theta _{\delta }& =0  \label{var1} \\
-\Delta \varphi _{\delta }& =\Delta \xi  \label{var2} \\
\theta _{\delta }+\lambda u_{\delta }+q\xi u_{\delta }& =0.  \label{var3}
\end{align}%
From (\ref{var2}) we get $\xi =-\varphi _{\delta },$ so (\ref{var1}) and (%
\ref{var3}) become%
\begin{align*}
-\Delta u_{\delta }+W^{\prime }(u_{\delta })+\theta _{\delta }(\lambda
-q\varphi _{\delta })& =0 \\
\left( \lambda -q\varphi _{\delta }\right) u_{\delta }& =-\theta _{\delta }.
\end{align*}%
From the above equations we clearly get (\ref{stat1}). (\ref{stat2}) is
given by the constraint (\ref{vic1}).

$\square $


\bigskip

\begin{thebibliography}{99}
\bibitem{azzpomp} \textsc{A.Azzollini, A.Pomponio}, \textit{Ground state
solutions for the nonlinear Klein-Gordon-Maxwell equations}, Topol. Methods
Nonlinear Anal., \textbf{35}, (2010), 33--42.

\bibitem{azzpipomp} \textsc{A.Azzollini, L.Pisani, A.Pomponio}, \textit{%
Improved estimates and a limit case for the electrostatic
Klein-Gordon-Maxwell system}, Proc. Roy. Soc. Edinburgh Sect. A, \textbf{141}%
, (2011), 449-463.

\bibitem{BBR07} \textsc{M.Badiale, V.Benci, S.Rolando,} \emph{A nonlinear
elliptic equation with singular potential and applications to nonlinear
field equations}, J. Eur. Math. Soc., \textbf{9} (2007), 355--381

\bibitem{BBBM} \textsc{J. Bellazzini, V. Benci, C. Bonanno, A.M.\ Micheletti,%
} \emph{\ Solitons for the Nonlinear Klein-Gordon-Equation}, Advances in
nonlinear studies, \textbf{10} (2010), 481-500.

\bibitem{bbbs} \textsc{J. Bellazzini, V. Benci, C. Bonanno, E. Sinibaldi, }%
\textit{Hylomorphic solitons in the nonlinear Klein-Gordon equation, }%
Dynamics in partial differential equations, \textbf{6} (2009), 311-333.

\bibitem{benci} \textsc{V. Benci, }\textit{Hylomorphic solitons, }Milan J.
Math., \textbf{77} (2009), 271-332.

\bibitem{bebo12} \textsc{V. Benci, C. Bonanno} \textit{Solitary waves and
vortices in non-Abelian gauge theories with matter, }Advanced Nonlinear
Studies \textbf{12} (2012), 717--735, arXiv:1105.5252

\bibitem{BF02} \textsc{V. Benci, D. Fortunato,}\ \textit{Solitary waves of
the nonlinear Klein-Gordon field equation coupled with the Maxwell
equations, }Rev. Math. Phys. \textbf{14} (2002), 409-420.

\bibitem{befogranas} \textsc{V. Benci, D. Fortunato,}\textit{\ Solitary
waves in the Nolinear Wave equation and in Gauge Theories, }Journal of fixed
point theory and Applications, \textbf{1}, n.1 (2007), 61-86.

\bibitem{befo} \textsc{V. Benci, D. Fortunato,}\textit{\ Solitary waves in
Abelian Gauge Theories, }Adv. Nonlinear Stud. \textbf{3 }(2008), 327-352.

\bibitem{befo08} \textsc{V. Benci, D. Fortunato,}\textit{\ Existence of
hylomorphic solitary waves in Klein-Gordon and in Klein-Gordon-Maxwell
equations,} Rend. Lincei Mat. Appl. supplemento, \textbf{20 }(2009), 243-279.

\bibitem{befoni} \textsc{V. Benci, D. Fortunato, }\textit{Hylomorphic
solitons on lattices, }Discrete and continuous Dynamical system, \textbf{28}
(2010), 875-897.

\bibitem{befospin} \textsc{V. Benci, D. Fortunato, }\textit{Spinning Q-balls
for the Klein-Gordon-Maxwell equations}, Commun. Math. Phys., 295 (2010)
639-668, doi: 10.1007/s00220-010-0985-z

\bibitem{befo11} \textsc{V. Benci, D. Fortunato, }\textit{Hamiltonian
formulation of the Klein-Gordom-Maxwell equations, }Rend. Lincei Mat.Appl. 
\textbf{22 }(2011), 1-22.

\bibitem{befo11max} \textsc{V. Benci, D. Fortunato}, \textit{On the
existence of stable charged Q-balls}, J. Math. Phys. 52, (2011), doi:
10.1063/1.3629848

\bibitem{befolak} \textsc{V. Benci, D. Fortunato, }\textit{A minimization
method and applications to the study of solitons, }Nonlinear Analysis,
T.M.A. \textbf{75} (2012), 4398---4421.

\bibitem{befolib} \textsc{V. Benci, D. Fortunato, }\textit{Hylomorphic
Solitons, }book\textit{\ }in preparation.

\bibitem{BL81} \textsc{H.Berestycki, P.L. Lions,} \textit{Nonlinear Scalar
Field Equations, I - Existence of a Ground State}, Arch. Rat. Mech. Anal., 
\textbf{82} (4) (1983), 313-345.

\bibitem{ca} \textsc{D. Cassani, }\textit{Existence and non-existence of
solitary waves for the critical Klein-Gordon equation coupled with Maxwell's
equations, }Nonlinear Anal.\textbf{58 }(2004), 733-747.

\bibitem{Coleman86} \textsc{S.Coleman,} \emph{\textquotedblleft
Q-Balls\textquotedblright }, Nucl. Phys. \textbf{B262} (1985), 263--283;
erratum: \textbf{B269} (1986), 744--745.

\bibitem{tea} \textsc{T.D'Aprile, D.Mugnai ,}\textit{\ Solitary waves for
nonlinear Klein-Gordon-Maxwell and Schr\"{o}dinger -Maxwell equations, }%
Proc. of Royal Soc. of Edinburgh, section A Mathematics, \textbf{134 }%
(2004), 893-906.

\bibitem{tea2} \textsc{T.D'Aprile, D.Mugnai,}\textit{\ \ Non-existence
results for the coupled Klein-Gordon- Maxwell equations, }Advanced Nonlinear
studies, \textbf{4} (2004), 307-322.

\bibitem{dav} \textsc{P.D'Avenia, L.Pisani, }\textit{Nonlinear Klein-Gordon
equations coupled with Born-Infeld Equations \ }Electronics J. Differential
Equations, \textbf{26} (2002), 1-13.

\bibitem{der} \textsc{G.H. Derrick, }\textit{Comments on nonlinear wave
equations as models for elementary particles , }J. Math. Phys. \textbf{5}
(1964), 1252-1254.

\bibitem{dod} \textsc{S. Dodelson, L. Widrow}, \textit{Baryon Symmetric
Baryogenesis}. Physical Review Letters \textbf{64}, (1990), 340--343.

\bibitem{em} \textsc{D. Eardley, V. Moncrief, }\textit{The global existence
of Yang-Mills-Higgs fields in }$\mathbf{R}^{3+1},$ Comm. Math. Phys. \textbf{%
83 }(1982), 171-212.

\bibitem{en} \textsc{K. Enqvist, J. McDonald}, \textit{Q-Balls and
Baryogenesis in the MSSM}, Physics Letters B \textbf{425,} (1998) 309--321.
arXiv:hep-ph/9711514.

\bibitem{Gelfand} \textsc{I.M. Gelfand, I.M. Fomin .}, \textit{Calculus of
Variations}, Prentice-Hall, Englewood Cliffs, N.J. 1963.

\bibitem{kus} \textsc{A. Kusenko, M. Shaposhnikov,} \textit{Supersymmetric Q
balls as dark matter}, Physics Letters B \textbf{418}, (1998), 46--54.
arXiv:hep-ph/9709492

\bibitem{leepa} \textsc{T. D. Lee, Y. Pang,} \textit{Nontopological solitons}%
, Physics Reports 221, (1992), 251--350.

\bibitem{long06} \textsc{E .Long, }\textit{\textit{Existence and stability
of solitary waves in non-linear Klein-Gordon-Maxwell equations, }}Rev. Math.
Phys. \textbf{18} (2006), 747-779.

\bibitem{kleinerman} S. \textsc{Klainerman, M: Machedon, }\textit{On the
Maxwell-Klein-Gordon equation with finite energy, }Duke Math. J. \textbf{74 }%
(1994), 19-44.

\bibitem{mug} \textsc{D. Mugnai, }\textit{Solitary waves in Abelian Gauge
Theories with strongly nonlinear potentials} Ann. Inst. H. Poincar\'{e}
Anal. Non Lin\'{e}aire \textbf{27} (2010), 1055-1071.

\bibitem{petre} \textsc{D.M. Petrescu , }\textit{Time decay of solutions of
coupled Maxwell-Klein-Gordon equations, }Coomm. Math. Phys., \textbf{179}
(1996), 11-24.

\bibitem{Rajaraman} \textsc{R. Rajaraman}, \textit{Solitons and instantons},
North Holland, Amsterdam, Oxford, New York, Tokio, 1988.

\bibitem{rosen68} \textsc{G. Rosen}, \emph{Particle-like solutions to
nonlinear complex scalar field theories with positive-definite energy
densities}, J. Math. Phys. \textbf{9} (1968), 996--998

\bibitem{rub} \textsc{V. Rubakov., }\textit{Classical theory of Gauge
fields, }Princeton University press, Princeton 2002.

\bibitem{yangL} \textsc{Y. Yang,} \textit{Solitons in Field Theory and
Nonlinear Analysis, } Springer, New York, Berlin, 2000.
\end{thebibliography}
\end{document}